\documentstyle[11pt,aaspp4]{article}

\def\gapprox{
        \mathrel{
                \vcenter{
                        \offinterlineskip \hbox{$>$}
                        \kern 0.3ex \hbox{$\sim$}
                }
        }
}
\def\lapprox{
        \mathrel{
                \vcenter{
                        \offinterlineskip \hbox{$<$}
                        \kern 0.3ex \hbox{$\sim$}
                }
        }
}

\newcommand{\eq}{\begin{equation}}
\newcommand{\ee}{\end{equation}}

\begin{document}

\title {On the Two-Phase Structure of Protogalactic Clouds$^1$}

\author {D. N. C. Lin\altaffilmark{2} \& S. D. Murray\altaffilmark{3}}

\bigskip
\bigskip
\bigskip
\bigskip
\bigskip
\bigskip
\bigskip
\bigskip
\begin{flushleft}
$^1${Lick Observatory Bulletin No.}
\end{flushleft}
\altaffiltext{2}{UCO/Lick Observatory, University of California, Santa Cruz,
CA, 95064\\Electronic mail: lin@ucolick.org}
\altaffiltext{3}{Lawrence Livermore National Laboratory, L-22, P.O. Box 808,
Livermore, CA, 94550\\Electronic mail: sdmurray@llnl.gov}

\begin{abstract}
In the gaseous envelope of protogalaxies, thermal instability leads to
the formation of a population of cool fragments which are confined by
the pressure of a residual hot background medium. In order to remain
in a quasi-hydrostatic equilibrium, the residual gas evolves at
approximately the virial temperature of the dark matter halo.  Its
density is determined by the requirements of thermal equilibrium.
The hot gas is heated by compression and shock dissipation.  The
heating is balanced by direct energy loss due to bremsstrahlung emission,
and by conductive losses into the cool clouds, which are efficient
radiators.  The cool fragments are photoionized and heated by the
extragalactic UV background and nearby massive stars.  Several processes
interact to determine the size distribution of the cool fragments.  The
smallest are evaporated due to conductive heat transfer from the hot gas.
All fragments are subject to disruption due to hydrodynamic instabilities.
The fragments also gain mass due to collisions and mergers, and condensation
from the hot gas due to conduction.  The size distribution of the
fragments in term determines the rate and efficiency of star formation
during the early phase of galactic evolution.  We have performed
one-dimensional hydrodynamic simulations of the evolution of the hot
and cool gas.  The cool clouds are assumed to follow a power-law
size distribution, and fall into the galactic potential, subject to drag
from the hot gas.  The relative amounts of the hot and cool gas is 
determined by the processes discussed above, and star formation occurs
at a rate sufficient to maintain the cool clouds at 10$^4$~K.  We present
density distributions for the two phases and also for the stars for
several cases, parametrized by the circular speeds of the potentials.
Under some conditions,
primarily low densities of the hot gas, conduction is more efficient than
radiative processes at cooling the hot gas, limiting the x-ray radiation
from the halo gas.

\end{abstract}

\section{Introduction}
According to CDM models of galaxy formation (Blumenthal et al. 1984;
Navarro, Frenk, \& White 1997; Klypin, Nolthenius, \& Primack 1997),
large galaxies
emerge through the coalescence of smaller systems, normally
identified as dwarf galaxies, which contain mostly dark matter with a
fraction of gas. The dark matter components merge with each other in a
dissipationless manner to form the halos of larger galaxies.  The
crossing of gas streamlines leads to shocks, heating the gas to the
virial temperature of the galactic halo (Binney 1977; Rees \& Ostriker
1977; White \& Rees 1978).  In order for the heated gas in this
extended halo (hereafter, the protogalactic cloud or PGC) to collapse,
its cooling timescale $\tau_c$ must be shorter than its dynamical
time scale, $\tau_d$ during each growth stage of the galaxy.  When
large galaxies acquire a mass comparable to the Galaxy, this condition
is satisfied once the characteristic length scale of the PGC is $<
100$ kpc (Blumenthal et al. 1984).

For typical values ($\sim 10^6$ K) of the virial temperature, the
cooling timescale increases with temperature, and the PGC's are
thermally unstable (Field 1965). Thermal instability leads to the
rapid growth of perturbations and fragmentation of PGCs (Murray \& Lin
1990). The result is that a two-phase medium develops during the
initial cooling of the PGC, in which a population of warm fragmentary
clouds (WFC's) are confined by the pressure of hot, residual halo gas
(RHG) (Burkert \& Lin 2000).  The RHG is primarily heated by the
release of the gravitational energy of the collapsing PGC and cooled
by radiative emission and conductive transport.  The WFC's settle into
the central region of the halo potential.  They are unable to cool below
10$^4$~K until their density reaches a sufficiently high value that
the WFC's become self-shielded from external photodissociating
UV radiation (Couchman \& Rees 1986; Haiman, Rees, \& Loeb 1997; 
Haiman, Abel, \& Rees in press).  Thereafter, molecular hydrogen can
form within them as
a consequence of non-equilibrium cooling.  Efficient cooling reduces
the temperature to $T < 100$~K even in the absence of metals.
Consequently, WFC's evolve into cold molecular clouds (CMCs).  The
formation of massive stars within the CMC's provides ultraviolet
heating, limiting the formation rate of CMC's from WFC's, and leading
to the formation of an equilibrium three-phase structure in the
galaxy.  In this paper, we discuss the evolution of the hot phase that
is left after the initial thermal instability.  Detailed
investigations of the warm and cold phases will be made in upcoming
papers.  For the current work, we assume a power-law size distribution
for the WFC's. The CMC's are not included explicitly, but star
formation within them is explicitly assumed to occur at a rate which
would provide an adequate flux of UV photons to heat and ionize the
WFC's.

In \S2, we briefly recapitulate the physical processes associated with
this scenario.  In \S~3, we determine the energy balance between the
warm and hot phases, while in \S~4, we discuss the resulting star
formation rate in the WFC's.  In \S~5, we show model results on the
evolution of the warm and hot phases.  Finally, we summarize our
results and discuss their implications in \S~6.

\section{The Emergence of a Two-phase Medium}
The temperature of a large, virialized PGC, $T \sim V_c^2 \mu/ k
\simeq 10^6 (V_c/100 {\ \rm km~sec}^{-1})^2 $~K, where $V_c$ is the circular
velocity of the halo potential, $\mu$ is the mean atomic weight in
grams, and $k$ is the Boltzmann constant. At such temperatures, the
dominant radiative cooling mechanisms are bremsstrahlung and
recombination processes (Dalgarno \& McCray 1972). The radiative cooling
rates for metal poor clouds (with [Fe/H]$\lapprox -2$) are
\begin{equation}
\Lambda(T)\approx10^{-22}\left(T\over{10^5~{\rm K}}\right)^{-1.2} \
{\rm ergs}\ {\rm cm}^{3}\ {\rm s}^{-1}
\end{equation}
for $2\times10^4$~K $< T<10^6$~K, and
\begin{equation}
\Lambda(T)\approx6\times10^{-24}\left(T\over{10^6~{\rm K}}\right)^{1/2}
\ {\rm ergs}\ {\rm cm}^{3}\ {\rm s}^{-1}
\end{equation}
for $T\gapprox10^6$~K.  Equation~(1) is a highly simplified average
representation of the behavior of the cooling rate in the given temperature
range.  In actuality, there is a peak due to H emission at
$2\times10^4$~K, another due to emission by He$^+$ at 10$^5$~K, and 
a minimum at $\approx10^6$~K, before the cooling efficiency rises again,
due to bremsstrahlung emission (Gould \& Thakur 1970).

The cooling timescale is
\begin{equation}
\tau_c={{{3\over2}\rho kT}\over{\mu n^2\Lambda}}
\end{equation}
where $\rho$ and $n$ are the mass and number density of the gas.  The
necessary collapse condition, $\tau_c\leq\tau_d$ where the dynamical
timescale $\tau_d=R/V_c$ (White \& Rees 1978), for a virialized PGC is
satisfied when $n$ exceeds a critical value
\begin{equation}
n_{crit}\approx0.2\ R_{kpc}^{-1}\left({T\over{10^6{\ \rm K}}}\right)^{1\over2}
{\ \rm cm}^{-3},
\end{equation}
where $R_{kpc}=R/1$~kpc, and R is the galactocentric radius.  

Above $3\times10^4$~K, $\tau_c$ is an increasing function of $T$, such
that the contrast between slightly cooler regions and the background
grows during the cooling process. The relatively rapid loss of entropy
also leads to a deficit of pressure, $P$, in the clouds compared to
their background.  The higher pressure of the background compresses
the clouds in an attempt to maintain pressure balance.  The rise in
$n$ leads to a reduction in $\tau_c$ for the clouds, further enhancing
the dichotomy in cooling timescale between the clouds and the
background.  The interface separating cool clouds from the background
retreats at an accelerating pace.  Although $\tau_c\sim\tau_d$ at the
onset of thermal runaway, it decreases rapidly to $\ll\tau_d$ as the
perturbed region cools significantly below the virial temperature.

Clouds, with radius $S$ and sound speed $c_s$, establish hydrostatic
equilibrium with the RHG and undergo isobaric evolution on a timescale
$\sim S/c_s$.  If this timescale is $> \tau_c$, $n$ cannot adjust such
that cooling is isochoric.  Relatively large clouds undergo a
transition from isobaric to isochoric cooling before their density
becomes significantly larger than that of the background.  This
transition in small clouds occurs at a later cooling stage.  Their
density growth becomes nonlinear before the cooling becomes isochoric.
Consequently, they emerge as dense, cool fragments (Burkert \& Lin
2000).

When $T$ in the clouds decreases to $\sim 10^4$K, recombination
becomes the dominant cooling process.  However, these clouds are also
exposed to UV radiation due to emission from background active
galactic nuclei.  At redshift $z= 2.5$, the flux of this emission at
the Lyman limit is estimated to be
$F_{UV}=0.5\times10^{-21}$~ergs~Hz$^{-1}$~ster
$^{-1}$~cm$^{-2}$~s$^{-1}$, with a dropoff towards higher energies as
steep as $\nu^{-3}$ (Haardt \& Madau 1996).  The above value varies
approximately as $(1+z)^3$.  In the outer regions of PGC where the
pressure of the RHG is low, the internal density of these clouds is
also relatively small, such that they may be mostly photoionized, with
$T\sim 10^4$K. But, in the inner regions of the PGC where the
background pressure is higher, WFC's have larger internal density,
such that they become self shielded against the extragalactic UV flux
and their temperature cools below $100$~K due to emission by H$_2$ and
metal ions (Dalgarno \& McCray 1972; Hollenbach \& McKee 1979).  When
these clouds become gravitationally unstable, stars rapidly form
within them.  The UV flux provided by these massive stars is able to
heat gas within their Str\"omgren radii to $10^4$~K, quenching star
formation in those regions.

At all galactocentric radii, then, the PGC fragments into a population
of WFC's which are maintained at $10^4$~K by the UV flux.  In most
regions, the ionizing photons are emitted by a population of massive
stars whose formation rate is self-regulated.  The WFC's are embedded
within RHG, whose density is sufficiently low that $\tau_c>\tau_d$,
ie. it is thermally stable (Field 1965), and it can remain near the
virial temperature of the galaxy.  The pressure of the RHG confines
all but the largest WFC's, for which self gravity is important.
Consequently, the PGC becomes a two-phase medium, with the density
contrast between the WFC's and the RHG inversely proportional to their
temperature ratio ($\sim100$).

\section{Energy Budget of the Two-Phase Medium}

\subsection{Dynamical Interaction Between the Warm and Hot Phases}

Following its collapse into the potential of the galactic halo, the
RHG is shock-heated to the virial temperature of the potential, and
rapidly attains a quasistatic equilibrium.  It then adjusts to have
density and temperature appropriate for the maintenance of thermal
equilibrium constrained by the processes discussed below.  The inverse
buoyancy of the pressure-confined WFC's causes them to settle towards
the center of the potential provided by the dark matter halo. During
their descent, the WFC's experience a drag by the RHG, reaching a
terminal speed relative to the RHG given by
\begin{equation}
V_t \approx \left(D_{\rho} {S\over R}\right)^{1/2} V_c,
\end{equation}
where $D_{\rho}\equiv\rho_w/\rho_h$ is the ratio of the density of the
WFC's to that of the RHGs.  (In all relations below, the subscripts
``h'' and ``w'' refer to quantities of the RHG and of the WFC,
respectively.)  For WFC's with sizes $S > R/D_{\rho}$, $V_t \sim V_c$.
Here, we neglect the effect of the clouds' size evolution on their
kinematics.

The value of $V_t$ derived in equation~(5) assumes a constant density
background.  In reality, the pressure gradient of the background causes
the density of the RHG to rise towards the center of the galaxy.  The
radii of the WFC's $S\ll R$, and so the pressure gradient across an
individual cloud is negligible.  If the RHG is approximately isothermal,
and the temperature of the WFC's are determined by atomic cooling to be
$\approx10^4$~K, then $D_\rho$ remains constant during the motion of
the clouds.  As a result, the primary effect of the density gradient of
the RHG upon $V_t$ is in altering the ratio $S/R$.  If
$\rho_h\propto R^{-a}$, then, under the above assumptions,
$\rho_w\propto R^{-a}$, $S\propto R^{a/3}$, and
$V_t\propto R^{(a-6)/6}$.

The motion of the WFC's through the RHG also leads to mass loss from
the WFC's due to
Kelvin-Helmholtz instability (Murray et al. 1993). In the limit
$D_{\rho}\gg1$, the growth timescale of KH instability is given by
\begin{equation}
\tau_{KH}=\lambda{{(\rho_h+\rho_w)}\over{(\rho_h\rho_w)^{1/2}V_t}}
\approx{{\lambda D_{\rho}^{1/2}}\over V_t}.
\end{equation}
Short wavelength perturbations rapidly saturate.  Perturbations with
wavelengths $\lambda\approx S$ provide the dominant contribution to
mass loss, which occurs at a rate
\begin{equation}
\dot M \simeq {4 \pi \rho_w S^2 \lambda \over \tau_{KH}}
\approx{4 \pi \rho_w S^2 V_t \over {D_{\rho}^{1\over2}}}.
\end{equation}
For a WFC with a mass $M = 4 \pi \rho_w S^3/ 3$, the stripping
timescale $\tau_s = M/ \dot M \sim (R/ 3 V_c) (S/R)^{1/2} < \tau_d$,
{\it i.e.} WFC's are disintegrated before they settle to the center of
the galactic potential.  Thus, both thermal instability associated
with the cooling, and Kelvin-Helmholtz instability arising during the
infall of the WFC's leads to the formation of WFC's with arbitrarily
small sizes.

The break down process increases the collective area filling factor of
the WFC's, leading to an increase in their collision frequency.
Collisions and the resulting mergers increase the masses of the WFC's.
When the rates of coagulation and disruption (including other
processes discussed below) balance with each other, an equilibrium
size distribution is established (Dong et al. in preparation).

\subsection{Conductive Heat Transfer Between Warm Fragments and Hot Medium}

The existence of WFC's in the RHG also leads to heat transfer from the
RHG to the WFC's through conduction (McKee \& Cowie 1977).  This
process is particularly important for small WFC's.  The conductive
heat flux into a WFC is given by
\begin{equation}
F_c = \kappa{\bf \nabla}T\ {\rm ergs}\ {\rm cm}^{-2}\ {\rm s}^{-1},
\end{equation}
where $\kappa\approx6\times10^{-7}T^{5/2}$ (Spitzer 1956).  To 
order of magnitude,
\begin{equation}
F_c\sim\kappa {T\over l}\approx\left({1\over3}\kappa_h T_h n_w ^2
\Lambda_w\right)^{1\over2},
\end{equation}
where $l$ is estimated as the depth into the WFC's over which
radiative emission balances conduction into the WFC's, and $\Lambda_w$
is the cooling efficiency within the WFC.  In terms of the density
contrast,
\begin{equation}
F_c\sim{{\mu_h}\over{\mu_w}}D_\rho\left({1\over3}\kappa_h T_h
\Lambda_w\right)^{1\over2}n_h,
\end{equation}
where $\mu_h$ and $\mu_w$ are the mean molecular weights in the RHG and
WFC's, respectively, and $n_h=\rho_h/\mu_h$.  In small clouds where the
density scale height is small, the magnitude of $F_c$ is limited by the
saturated flux,
\begin{equation}
F_c \simeq F_{sat} = \rho_h c_h^3,
\end{equation}
where $c_h$ is the isothermal sound speed in the RHG
($\approx$~220~km~s $^{-1}$ for the Galaxy).  The presence of magnetic
field reduces the mean free path of the electrons and $F_c$ (Rosner \&
Tucker 1989).
In our calculations, we neglect the effect of magnetic field and adopt
$F_c = F_{sat}$ based on our model parameters.

Large WFC's, which attain large $V_t$, may also advect thermal energy from 
the RHG to the WFCs at a rate
\begin{equation}
F_a \approx {1\over2}\rho_h V^3.
\end{equation}
As discussed above, the sizes of the WFC's are modified by
Kelvin-Helmholtz instability and coagulation before they are
accelerated to $V_t$.  The WFC's may therefore spend most of their
time with smaller velocities relative to the RHG (Murray et al. 1993).
The advective heating is therefore expected to be weaker than that due
to conduction.

\subsection{The Warm Fragmentary Clouds}
Energy equilibrium is established within a cloud when
$${4\over3}\pi S^3 \Gamma_r + 4\pi S^2 F_c = {4\over3}\pi S^3
n_w^2\Lambda_w.$$ The photoionization heating
\begin{equation}
\Gamma_r = (1-x_w)n_w h\int_{\nu_0}^{\infty}{{4\pi J_\nu}\over{h\nu}}
a_{\nu}(\nu-\nu_0)d\nu= (1-x_w) n_w f_G4\pi J_{\nu_0}a_{\nu_0}\nu_0,
\end{equation}
where $x_w$ is the ionization fraction in the WFC's, $h$ is Planck's
constant, $J_\nu$ is the mean intensity in
ergs~ster$^{-1}$~Hz$^{-1}$~s$^{-1}$, $a_\nu$ is the photoionization
cross section as a function of frequency, $h\nu_0=13.6$~eV, and $f_G$
depends upon the shape of the spectrum ($f_G<0.1$ for realistic
background spectra).  This equilibrium determines the temperature of
the WFC's.  The above assumes a top-hat density distribution within the 
clouds.  This is appropriate for clouds which are at most weakly
self-gravitating, given the sharp temperature transition between the 
WFC's and the RHG resulting from thermal instability.

In a mostly ionized WFC, the heat input due to conduction is larger
than that due to photoionization when
\begin{equation}
x_w > x_c = 1-{3\over S}{{F_c}\over{n_w f_G 4\pi J_{\nu_0}a_{\nu_0}\nu_0}}.
\end{equation}
In this limit, the rate of conductive heat input exceeds that of the
radiative loss due to recombination  when
\begin{equation}
S < S_{es} = {{3\mu_h n_h c_h ^3}\over{\Lambda_w n_w ^2}}
\end{equation}
in the limit of saturated conduction and
\begin{equation}
S < S_{eu} = \left({{3\kappa_h T_h}\over{n_w^2 \Lambda_w}}\right)^{1/2}
\end{equation}
in the limit of unsaturated conduction.  The above results also apply
to the limit in which the background UV flux is negligibly small, so
that the gas in WFC's is mostly neutral.  For the intermediate case in
which photoionization contributes significantly to the heating, the
additional energy input would tend to increase the value of $S_{eu}$
and $S_{es}$.

WFC's with $S<S_{es}$ (or $S_{eu}$) are unstable to conductive heat
transport.  Their inability to radiate away all the conductive heat
input leads to an increase in their temperatures.  In order to
maintain pressure equilibrium with the RHG, these WFC's expand,
decreasing their internal density.  The increase in $S$ enhances
conductive heat transport into the clouds, whereas their cooling
efficiency decreases with the decrease in $n_w$.  Unless $\Lambda_w$
increases rapidly with $T_w$, conduction then leads to the total
evaporation of these small WFC's.  The stability criterion can be
obtained by analyzing the dependence of $S_{es}$, $S_{eu}$, and $S$
upon $T_w$.  If we approximate $\Lambda_w \propto T_w^s$, then from
equations~15 and 16 we find that both $S_{es}$ and $S_{eu}$ vary as
$T_w^{(2-s)/2}$.  As the pressure-confined WFC's are heated and
expand, however, their radii vary as $T_w^{1/3}$.  The clouds are
therefore unstable if $s<4/3$.  In that case, we can picture a cloud
with $S=S_{es}$ initially.  As it is heated and expands, the steep
dependence of $S_{es}$ upon $T_w$ leads to $S<S_{es}$.  Radiative
cooling can then no longer offset conduction into the cloud, leading
to further temperature increase and expansion.  The same stability
criterion of $s>4/3$ holds for clouds where $S_{eu}$ is the
appropriate limiting radius.  This stability condition is met over the
temperature range dominated by hydrogen emission, $9000{\ \rm K}\leq
T_w\leq20,000$~K, such that clouds with $S<S_{es}$ or $S<S_{eu}$ 
remain stable until their temperatures reach 20,000~K.  The minimum
possible stable cloud size is therefore set by the value of $S_{es}$
or $S_{eu}$ at 20,000~K, where $\Lambda(T)$ is a maximum.

In the above analysis, the WFC's are assumed to be confined by the
ambient pressure of the RHG.  If the WFC's are able to cool to
$T\lapprox 10^2$~K, then self-gravity contributes to the confinement
of even relatively small clouds. In this limit, thermal equilibrium
between photoionization heating and radiative losses may still be
established, but it is unstable because of the negative specific heat
of self-gravitating clouds (Murray \& Lin 1992).  The result is
similar to that for non self-gravitating clouds, except that, for a
self-gravitating cloud, the radius varies as $T_w^{-1}$, and stability
requires $s>4$, which is, again, only satisfied over a very narrow
range of temperatures.

While the lower limit of cloud radii is set by their ability to
radiate away the energy flux due to conduction in equations~(15) and (16),
the upper limit of cloud radii is set by gravitational instability.
Low mass clouds are confined by the pressure of the RHG.  At higher
masses, self-gravity becomes important.  Clouds become unstable above
a critical mass, which depends upon the external pressure
\begin{equation}
M_{crit}=\left[3.15\left({{kT_w}\over{\mu_w}}\right)^4 {1\over{G^3 P_h}}
\right]^{1\over2},
\end{equation}
where $P_h$ is the pressure of the RHG, and $T_w$ is the internal temperature
of the clouds (Ebert 1955; Bonner 1956).  This condition corresponds to an 
approximate maximum cloud radius of
\begin{equation}
S_{BE}=1.2\left({k\over G}\right)^{1\over2}\left({3\over{4\pi}}\right)
^{1\over3}\left({{T_w}\over{\mu_w}}\right)\left(n_hT_h\right)^{-{1\over2}}.
\end{equation}
We have adopted the Bonner-Ebert criterion for stability, rather than
the Jeans criterion, due to the fact that the WFC's are pressure-confined.
While there might be an apparent inconsistency, in that we assume the
WFC's to have nearly constant density, whereas the Bonner-Ebert 
criterion assumes a density gradient.  Had we used the Jeans criterion,
however, our value for $S_{BE}$ would be very close to that used
above, and so the choice of stability criterion does not affect our
results below.

The above results assume that the WFC's are subject only to heating
by photoionization and by conduction from the hot gas.  In addition,
the drag of the clouds through the RHG, and the growth of Kelvin-Helmholtz
instability will both lead to additional mechanical energy input into the
WFC's which could, in principle, provide additional support for the
clouds.  These mechanisms are least important, however, in the largest
clouds, which
experience the least drag, and whose self-gravity stabilizes the growth of
deeply-penetrating, long wavelength perturbations by Kelvin-Helmholtz 
instability (Murray et al. 1993).  Stability of the most massive clouds
should therefore still be governed by the Bonner-Ebert criterion.
Detailed models of clouds moving through background gas will be used in
future work to confirm these conclusions.

\subsection{Total Rate of Conductive Energy Transport}

The physical processes involved in determining the mass spectrum of
the WFC's will be examined in a future paper.  In the results below,
we assume that the fragmentation and coagulation of the WFC's leads to
a power-law size distribution of the WFC's, in which the number
density of clouds within the PGC, $n_c$, per unit cloud radius $S$ is
given by
\begin{equation}
{{dn_c}\over{dS}} = A S^\gamma.
\end{equation}
The normalization constant, $A$, is related to the spatially averaged
density of the warm phase by
\begin{equation}
\langle\rho_w\rangle=\int{{4\over3}\pi S^3\rho_w
{ {{\rm d}n_c}\over{{\rm d}S} }{\rm d}S} = {4\over3}\pi\rho_w A
\left({1\over{\gamma+4}}\right)\left(S_{max}^{\gamma+4} -
S_{min}^{\gamma+4}\right)
\end{equation}
where $S_{max}$ and $S_{min}$ are the maximum and minimum cloud size,
respectively.  As discussed above, $S_{min}$ is determined by the
smallest size of clouds which can radiate away the energy conducted
into them from the RHG, while $S_{max}$ is determined by the
Bonner-Ebert mass.  The value of $A$ is a function of galactocentric
radius, whereas we assume $\gamma$ to be the same everywhere within
the PGC.

If the clouds follow a distribution close to that observed within
Galactic cloud complexes, for which $\gamma =-2.5 $ (Scalo 1985;
Zinnecker, McCaughrean, \& Wilking 1993 ), then the results above and
below are dominated by the massive end of the distribution.
Alternatively, if the efficiency of cloud disruption is high relative
to the rate of coagulation, then the cloud distribution will be
dominated by the low-mass end.  The determination of which limit is
appropriate shall be made in an upcoming paper.  In the current work,
we treat the size distribution as an uncertainty, and consider two
extreme limits.  In the first, we assume $\gamma>-3$, and all terms
derived both above and below are dominated by the high mass end of the
cloud distribution.  In the second, we assume $\gamma<-4.5$, and all
terms are dominated by the low mass end of the distribution.

The total rate of conductive energy transport per unit volume from the RHG
into the WFC's is given by
\begin{equation}
L_{cond} = \int {{dn_c}\over{dS}} F_c 4\pi S^2 dS.
\end{equation}
Using the results above, this becomes
\begin{equation}
L_{cond} = {{4\pi AF_c}\over{\gamma+3}}\left(S_{max}^{\gamma+3}-
S_{min}^{\gamma+3}\right)
\end{equation}
For $\gamma>-3$, the above result becomes
\begin{equation}
L_{cond}=3\left({{\gamma+4}\over{\gamma+3}}\right){{\langle\rho_w\rangle}
\over{\rho_w}}F_c S_{BE}^{-1},
\end{equation}
while for $\gamma<-4$ it is
\begin{equation}
L_{cond}=3\left({{\gamma+4}\over{\gamma+3}}\right) {{\langle\rho_w\rangle}
\over{\rho_w}} F_c S_{es}^{-1}.
\end{equation}
The above results assume that $S_{max}\gg S_{min},$ where $S_{max}=S_{BE}$
and $S_{min}=S_{es}$.  As can be seen by a
comparison of equations~(15) and (18),
$S_{min}/S_{max}\sim 0.01 (\Lambda_w/10^{-24})^{-1}n_h^{-1/2},$ and so
this is a reasonable approximation.  Also, because $S_{BE}\gg S_{es}$,
the greater surface-to-volume ratio of the case $\gamma<-4$ leads to
a significantly greater cooling rate due to conduction into the WFC's
relative to the case with $\gamma>-3$.

\subsection{The Residual Halo Gas}
Equations~(23) and (24) give the conductive energy transfer rate from
the RHG to the WFC's.  The RHG also experiences radiative losses,
mainly through bremsstrahlung emission, which occurs at a rate given
by equation~(2).  Using the conductive flux given in equations~(23) and
(24), we find that conductive losses into the WFC's exceed
bremsstrahlung cooling in the limit that
\begin{equation}
n_h^2<3 \left({{\gamma+4}\over{\gamma+3}}\right)
\left({{S_{max}^{\gamma+3}-S_{min}^{\gamma+3}}\over
{S_{max}^{\gamma+4}-S_{min}^{\gamma+4}}}\right){{\mu_h}\over{mu_w}}{{T_w}
\over{T_h}}{{c_h^3}\over{\Lambda_h}}\langle\rho_w\rangle.
\end{equation}
when conduction is saturated, and
\begin{equation}
n_h^2<\left({{\gamma+4}\over{\gamma+3}}\right)
\left({{S_{max}^{\gamma+3}-S_{min}^{\gamma+3}}\over
{S_{max}^{\gamma+4}-S_{min}^{\gamma+4}}}\right)
{{\left(3\kappa_h T_h \Lambda_w\right)^{1\over2}}\over{\Lambda_h}}
\langle\rho_w\rangle
\end{equation}
when conduction is unsaturated.  In the above relations, we assume
pressure balance between the warm and hot phases, ie. 
$n_h T_h = n_w T_w.$

Heat is supplied to the RHG through its own quasi-static
contraction. More importantly, heat is also deposited into the RHG as
a result of the frictional drag on collapsing WFC's, by which the
WFC's are able to transfer a fraction of their gravitational potential
energy into thermal energy of the RHG.  (In the absence of any drag,
the total energy of individual WFC's is approximately conserved as
they fall into the galactic halo potential.)  The heating per unit
volume of a flux of WFC's moving at their terminal speeds is given by
\begin{equation}
\Gamma = \int{ {{V_c^2}\over R} {{{\rm d}{n_c}}\over{{\rm d}S}}
{4\over3}\pi \rho_w S^3 V_t {\rm d}S}.
\end{equation}
Using $V_t$ from eq~(6), the above result becomes
\begin{equation}
\Gamma = \left({{\gamma+4}\over{\gamma+{9\over2}}}\right)
\left({{S_{max}^{\gamma+{9\over2}}-S_{min}^{\gamma+{9\over2}}}\over
{S_{max}^{\gamma+4}-S_{min}^{\gamma+4}}}\right)
\left({{\mu_wT_h}\over{\mu_hT_w}}\right){{V_c^3}\over{R^{3\over2}}}
\langle\rho_w\rangle.
\end{equation}

For $\gamma>-4$, the above result is dominated by the contribution of
the largest WFC's, for which $V_t$ may become comparable to $V_c$ at
small $R$.  Whether or not the clouds are able to attain $V_t$ depends
upon the ratio of their acceleration timescales to the timescales upon
which they are affected by conduction, collisions, or Kelvin-Helmholtz
instability.  For a cloud to be accelerated to a speed $\sim V_c$, it
must travel a distance $\sim R/2$ in the halo, whereas,
Kelvin-Helmholtz instability leads to significant mass loss by the
time a cloud has moved through the RHG over a distance of several
times its diameter (Murray et al. 1993).  Unless the RHG is also
contracting with a velocity $V_H \sim V_c$, the massive WFC's may not
be able to attain their terminal velocities (Dong et al. in
preparation).  Instead of $V_t$ above, we parametrize the terminal
differential velocity as $fV_c$, with $f$ being a free parameter.
Using this prescription, the heating rate is given by
\begin{equation}
\Gamma={{fV_c^3}\over R}\langle\rho_w\rangle.
\end{equation}

In the limit $\gamma<-4.5$, the heating is dominated by drag from low mass
clouds, for which $V_t<V_c$.  The above result then becomes
\begin{equation}
\Gamma = \left({{\gamma+4}\over{\gamma+{9\over2}}}\right)
{{V_c^3}\over{R^{3\over2}}}
\left({{\mu_wT_h}\over{\mu_hT_w}}\right)^{1\over2}\langle\rho_w\rangle
S_{es}^{1\over2}.
\end{equation}

If the radiative and conductive cooling of the RHG exceed the thermal
energy input, some of the RHG will precipitate to form additional
WFC's.  As the gas density of the RHG is depleted, both the conductive
heat flux into, and drag on the infalling WFC's decrease.  The density
$n_h$ therefore adjusts so as to achieve thermal equilibrium in which
$G \simeq L_{cond} + L_{brem}$.

\section{Ionization of the clouds}

\subsection{The Nature of the External Source}
In the above analysis, the WFC's are generally taken to be highly
ionized.  To maintain this ionization state, the UV flux to which the
WFC's are exposed must be sufficiently large to offset recombination.
In Figure~1, we show the depths into which a cloud may be ionized by
various sources of UV flux, plotted as a function of $n$.  These
results were obtained with CLOUDY (Ferland 1991).  The solid, dashed,
and dot-dashed curves represent results obtained with the standard AGN
spectral energy distribution, for three values of $F_{UV}$.  The solid
curve was calculated using a value of
$0.5\times10^{-21}$~ergs~Hz$^{-1}$~ster$^{-1}$~cm$^{-2}$~s$^{-1}$
at 13.6~eV,
representing the UV background at z=2.5 (Haardt \& Madau 1996).  The
dashed and dot-dashed curves were computed using fluxes that were,
respectively, smaller and larger by an order of magnitude,
approximately representing redshifts $<0.5$ and 5 respectively.  The
dotted curve was computed using the Kurucz spectrum in CLOUDY, with
the flux set to that of an O5 star at a distance of 10~pc.

From Figure~1, we can infer the integrated depth into the PGC outside
of which all of the warm fragments are ionized.  In the early evolutionary
epochs of the Galaxy before most of the ordinary matter was converted
into stars, only the regions outside $\sim 30$ kpc are photoionized by
the UV background (see below).  At smaller galactocentric radii, local
sources of UV radiation are required.  In diffuse and low mass dwarf
galaxies, however, the entire galaxy may be photoionized by the diffuse
background radiation (Kepner, Babul, \& Spergel 1997).

In addition to the extragalactic background, the RHG radiates soft
X-ray photons following a bremsstrahlung spectrum.  Previous work
(Fall \& Rees 1985), which used analytic approximations to estimate the
structure of the RHG, found the resulting x-ray flux to be
insufficient to ionize large clouds beyond a few kpc from the Galactic
center.  In the models presented below, we find the bremsstrahlung
luminosities to be less than estimated by Fall \& Rees (1985).
Ionization by the RHG is therefore not a significant contributor
to determining the state of the WFC's.

Hydrodynamic drag on WFC's also leads to energy input into them, as
described in \S~3.4.  For relatively large WFC's $V_t >
10$~km~s$^{-1}$.  If these fragments are not ionized initially, then
shocks will cause internal heating.  From above, though, the ratio of
heating by drag to that by conduction is found to be
\begin{equation}
{{{\pi\over2}\rho_1S^2V_t^3}\over{4\pi S^2F_c}}=3\times10^{-4}{\ }
\left({{V_t}\over{10{\ \rm km\ s}^{-1}}}\right)^{3}.
\end{equation}
Thus, even for velocities as high as 100~km~s$^{-1}$, the heating rate
due to drag is negligible. For survivable clouds with $S> S_{eu}$ and
$S_{es}$, conduction does not provide adequate heat transfer to offset
radiative loses.  Since the dominant radiative process at $\sim 10^4$K
is recombination, conductive heat flow is also insufficient to ionize
the WFC's.

\subsection{Self-Regulated Star Formation}
In the absence of an adequate ionizing photon flux, the cooling
efficiency of bremsstrahlung, recombination, and atomic hydrogen
emission decrease rapidly below $\sim 10^4$~K.  In a metal-free
protogalactic cloud, however, non-equilibrium recombination leads to
the formation of a small amount of $H^-$ ions which recombine with
neutral $H$ to form $H_2$.  Radiative losses due to rotational and
vibrational transitions of $H_2$ reduce the gas temperature to $\sim
10^2$~K (Murray \& Lin 1990).  If [Fe/H]$> -3$, even lower
temperatures ($\sim 10$ K) are attainable due to cooling by heavy
elements (primarily by CII, CO, and Silicate grains) (Hellsten \& Lin
2000), leading to the formation of cold molecular clouds (CMC's).

When the temperature of the CMC's is reduced by over two orders of
magnitude from that of the WFC's, their Bonner-Ebert masses are
reduced by over four orders of magnitude (cf. equation~17).
Self-gravity is therefore much more important for the evolution of the
CMC's than for that of the WFC's.  Cloudlets close to the Bonner-Ebert
mass are centrally condensed.  When their mass exceeds the
Bonner-Ebert mass, thermal pressure can no longer support the weight
of the envelope, and the cloudlets undergo inside-out collapse (Shu
1977).  During the collapse the Jean's mass decreases with
the increasing density.  Due to the centrally-concentrated cloud
structure, however, the Jeans mass is always larger than the mass
contained inside any given radius.  The collapse is therefore stable,
and does not lead to fragmentation in the absence of further unstable
cooling, such that the initial value of $M_{crit}$ represents the
minimum mass for isothermal
collapsing clouds (Tsai 1992). Numerical simulations show that even
rotating centrally-concentrated clouds have difficulties in breaking up
into fragments with a small fraction of $M_{crit}$ (Burkert \&
Bodenheimer 1996). This conjecture is in contrast to the conventional
opacity-limited fragmentation scenario (Hoyle 1953; Low \& Lynden-Bell
1976) which may be more appropriate for the collapse of unstable
homogeneous clouds (Burkert \& Bodenheimer 1993).  The initial conditions
required by the opacity-limited fragmentation scenario may be difficult
to accomplish in nature.

In a metal-free environment, the temperature of the CFC's,
$T\sim 10^2$~K and $M_{crit} \sim 10-10^2
M_\odot$.  Stars formed in such a metal-poor environment are massive
and short-lived, consistent with their rarity today (Hellsten \& Lin
2000).  The early chemical enrichment is dominated by the output
of Type II supernovae, consistent with the abundance distribution
observed among stars with [Fe/H]$<-1$ (Wheeler, Sneden, \& Truran
1989).

The massive stars are also copious sources of UV radiation.  The effect
is twofold.  Photons in the range 11-13.6~eV are able to penetrate
clouds and photodissociate H$_2$ within them, unless their column
densities are so high the clouds are self-shielded.  Higher energy
photons ionize and heat the gas to 10$^4$~K.  The loss of coolants,
and increased heating means that a
population of $\sim10^5$~O5 stars is adequate to photoionize all the
CMC's, re-heat them to $T\sim 10^4$~K, and prevent the formation of
any additional CMC's out to a few tens of kpc.  Photoionization
heating also increases $M_{crit}$ to $\sim 10^6 M_\odot$, stablizing
small cloudlets, and quenching star formation. As the massive stars
evolve off the main sequence, the UV flux decreases, cooling again
leads to $T\sim 10^2$~K in sheltered regions, and spontaneous star
formation is resumed.  Thus, the formation rate of massive stars may
be estimated from the assumption that it is self-regulated at a level
necessary to sustain marginal ionization of all WFC's.  This
self-regulated star formation rate naturally yields values of [Fe/H]
comparable to that observed in halo stars (Lin \& Murray 1992).

The rate at which massive stars must be formed is determined by the
need to balance ionization with recombination in the WFC's.  The mean
free path for Lyman continuum photons is large in the RHG, but not in
the WFC's.  For sufficiently high covering factor, we can assume that
a large fraction of the UV photons produced by hot stars are absorbed
locally.  In equilibrium then, the total production rate of UV photons
per unit volume must balance the total recombination rate per unit
volume.  This radiative equilibrium results in high ionization in the
limit
\begin{equation}
n_\ast Q_0 = n_w^2\alpha_B\int{ {4\over3}\pi S^3 {{{\rm d}n_c}\over{{\rm d}S}}
{\rm d}S}
=\left({{T_h}\over{T_w}}\right){{\alpha_B}\over{\mu_w}}\langle\rho_w\rangle,
\end{equation}
where $n_\ast$ is the number density of stars, $Q_0$ is the average UV
output in photons~s$^{-1}$ per star, and $\alpha_B$ is the case B
recombination coefficient (Osterbrock 1989).

The production of UV photons is dominated by O stars, with lifetimes
$\tau_\ast\approx3\times10^6$~yr.  The average mass loss rate per unit
volume from the warm phase due to star formation is then given by
\begin{equation}
\dot\rho_{SF}={{n_\ast m_{sf}}\over{\tau_\ast}}=
{{m_{sf}}\over{\tau_\ast}} {{\alpha_B}\over{Q_0}} {{T_h}\over{T_w}}
{{\langle\rho_w\rangle}\over{\mu_w}}n_h,
\end{equation}
where $m_{sf}$ is the mass of stars that form to produce $Q_0$ photons
per unit time.
If the stars follow a Salpeter mass function, with a minimum mass of
0.1~M$_\odot$ and maximum mass of 100~M$_\odot$, then 
\begin{equation}
{{Q_0}\over{m_{sf}}}\approx4\times10^{47}{\rm photons\ M}_{\odot}^{-1}
{\ \rm s}^{-1}.
\end{equation}

The rate of mass depletion by star formation in the warm phase derived
above is independent of the size distribution of the WFC's, and
depends only upon their filling factor and internal density.

\section{Protogalactic Models}

\subsection{The Numerical Scheme}

We have performed one-dimensional Lagrangian hydrodynamic models of
the evolution of the RHG subject to the physical processes discussed
above.  The models treat the warm and hot phases as separate fluids,
evolving within a fixed external gravitational potential set by the
dark matter.  Angular momentum of the gas is not included.  The
models are applicable to the very early stages of star formation in
galaxies or after mergers, or to cooling flows, before the gas has
contracted sufficiently for its motion is dominated by angular momentum.
During the later disk formation phase of a galaxy, or in the innermost
regions of mergers or cooling flows, angular momentum would play an
important role, whereas for the situations modelled here it would
not strongly affect the physical processes under consideration.

The momentum equation for the RHC  
\begin{equation}
{{{\rm d}u_h}\over{{\rm d} t}}=-4\pi R^2{{\partial P_h}
\over{\partial m}}-g-F_{drag},
\end{equation}
includes contributions from the gas pressure, $P_h$, the dark matter
potential, $g$, and drag from the infalling WFC's.  In equation~(35),
$u_h$ is the radial velocity of the RHG, and $4\pi\rho_h R^2{\rm
d}R={\rm d}m$.  For computational simplicity, the gravitational
potential is taken to be isothermal out to a radius of 50~kpc, such
that $g=V_c^2/R$.  The magnitude of $g$ is held constant within a core
radius $R_{core}=1$~kpc and dark matter is assumed to vanish outside
of 50~kpc, such that $g$ becomes Keplerian.  Although this
prescription of $g$ is highly simplified, our results do not depend
sensitively upon the detailed form of $g$.  The form of the drag term
is
\begin{equation}
F_{drag}=\int{\pi S^2\left(V_t-u_h\right)^2{{ {\rm d} n_c}\over
{ {\rm d} S}}{\rm d}S}.
\end{equation}
As noted above, in models where high mass clouds dominate, we set
$V_t=V_c$, whereas equation~(5) is used in models where low mass clouds
dominate.

The equation of continuity for the RHG,
\begin{equation}
{{{\rm d}\rho_h}\over{{\rm d}t}}=-\left[{{\rho_h}\over{R^2}}{\partial\over
{\partial R}}\left(R^2 u_h\right)+\dot\rho_c(R,t)\right],
\end{equation}
includes a sink term, due to condensation of gas out of the hot phase.
The amount of mass lost from the hot phase depends upon the relative
magnitudes of heating and cooling. If heating either exceeds or equals
cooling in the RHG, then $\dot\rho_c=0$.  If cooling exceeds heating
in the RHG, however, mass is transferred from the RHG to the WFC's at
a rate given by
\begin{equation}
\dot\rho_c={{\rho_h-\rho_{bal}}\over{\tau_{c}}},
\end{equation}
where $\rho_{bal}$ is the density which would be required for heating
to balance cooling in the hot phase.  The energy equation for the RHG,
\begin{equation}
{{{\rm d}e_h}\over{{\rm d} t}}={{\Gamma - \Lambda_{brem} -
\Lambda_{cond}}\over{\rho_h}},
\end{equation}
includes heating by drag from the warm phase, bremsstrahlung emission, and
conduction losses into the warm phase.

The warm phase is not pressure-supported in the galactic halo, and so
the mean velocity of the WFC evolves as
\begin{equation}
{{{\rm d}u_w}\over{{\rm d}t}}=-g+F_{drag},
\end{equation}
where the drag term is similar to that in equation~(35).  A maximum
infall velocity of $fV_c$ is imposed upon the warm phase.  As
discussed above, the parameter $f$ depends upon the details of the
evolution of the warm phase.  In our models, we consider values of
$f=1$ and 0.1, thus examining the differences in our results for
widely varying values of $f$.

The mass of the warm phase increases by mass loss from the hot phase,
and decreases by mass loss due to star formation, following the
prescription given in equation~(33), such that
\begin{equation}
{{{\rm d}\langle\rho_w\rangle}\over{{\rm d}t}}=
\dot\rho_c(R)-\left[{{\langle\rho_w\rangle}\over{R^2}}
{\partial\over{\partial R}}\left(R^2\langle\rho_w\rangle\right) +
\dot\rho_{SF}\right].
\end{equation}
In principle, the smallest WFC's (with $S < S_{eu}$ or $S < S_{es}$)
are continually being evaporated into the RHG.  But, if this phase
transition increases $\rho_h$ such that cooling exceeds heating in the
RHG, then an appropriate amount of hot gas would precipitate to retain
thermal equilibrium in the RHG.  Thus, the ratio $ <\rho_w> / \rho_h$
depends indirectly on the size distribution of WFC's which determines
the conductive flux between the two phases.

Finally, the WFC's are assumed to be ionized and heated by the UV flux
of the nearby massive stars, such that $T_w \simeq 10^4$ K.  They are
also assumed to be in a pressure equilibrium with the RGH such that
$n_w = n_h T_h/T_w$.  Equations~(35)-(41) completely describe the
$\rho_h$, $T_h$, $V_h$, $V_w$, and $< \rho_w >$ distribution.  From
these quantities, $T_w$, $\rho_w$, as well as the heat transfer
luminosity due to bremstrahlung and conduction, $L_{\rm brem}$ and
$L_{\rm cond}$ can be obtained.  We solve these equations numerically
with a 1-D Lagrangian method (Richtmyer \& Morton 1995).

The warm and hot phases are treated as separate fluids, evolving and
interacting as described above.  The models described below use a grid
of 400 cells for both fluids, with the cell spacing increasing
outwards so as to give the best resolution in the region of interest,
within 100~kpc.  The outer boundary is set at 2~Mpc, with a constant
pressure.  The large outer radius is chosen so that conditions at
that boundary cannot affect the solution.
Variations by factors of two in the number of cells are found to have
no significant effect upon the results.  The hydrodynamic equations
are evolved
explicitly, while the energy equation (38) is solved implicitly,
allowing the use of a simple Courant timestep criterion.  The models
are evolved until they reach a quasi-steady state, the results
of which are shown in the following sections.

\subsection{Model~1}
Using the approach outlined above, we now consider several simple
models to describe
the evolution of gaseous ordinary matter in the halo of large
galaxies.  In principle, we should concurrently consider the
cosmological evolution which led to the formation of the halo.
However, such an investigation would require not only a large
dynamical range in dissipationless interaction among dark matter
structure but also the nonlinear evolution of ordinary matter in
evolving and asymmetrical potentials.  A number of authors have
performed simulations of systems which follow the evolution of both
dissipational and dissipationless components, either in the context
of galaxy collisions, or galaxy formation (see, eg. Barnes \&
Hernquist 1991; Gerber, Lamb, \& Balsara 1996; Lamb et al. 1997;
Abel et al. 1998; Steinmetz \& Navarro 1997, 1999; and references therein).
Such studies are generally either unable to resolve the full dynamical
range present in a young galaxy, or do not include the full range of
physical processes discussed above.  Fortunately, we are able to
examine the importance of the processes discussed in \S~3 and 4 by
following the standard approach in galaxy formation simulations by
breaking the problem into piecemeal tasks.  In contrast to the conventional
simulations (eg. Yepes et al. 1997; Elizondo et al. 1999), which mainly
deal with the emergence of dark
matter structure and in which the star formation process is parameterized,
we assume the potential is already formed and focus our attention on the
microphysics of the gas.

In the first model, we take $V_c=220$~km~s$^{-1}$, and assume that
$\gamma>-3$, so that the appropriate equations to use are those
derived for limit in which high mass clouds dominate.  For these massive
clouds, the gas drag effect is relatively weak and we initially
adopt $f=1$, {\it i.e.} the WFC's move inward at speeds of $V_c$.

The initial conditions of the model are uniform density in the hot
phase, with $n_h=0.01$~cm$^{-3}$, giving a mass of
$\sim10^{11}$~M$_\odot$ within a radius of 50~kpc.  The average
density of the warm phase is initially taken to be 0.01 times that of
the hot phase.  The temperature of the hot phase is initially set to
10$^6$~K.  We have varied the initial conditions substantially with
little change in the final distribution of the hot phase.

The model was evolved for 10$^9$~yr, by which time it had reached an
asymptotic quasi-stationary state.  While the mass of gas continues to
decrease due to star formation, the density distributions of the RHC and
WFC's attain asymptotic forms.  These are shown in Figure~2.
Figure~2a shows the final radial distributions of the density of the
hot phase (solid curve), the average density of the warm phase (dashed
curve), and the density of stars formed from the warm phase during the
evolution (dot-dashed curve).  The stellar density is calculated using
the positions of stars where they are first formed.  Because the stars
will form on highly eccentric orbits, for which they spend most of their
time near apocenter, the calculated stellar distribution should be
representative of the observed distribution, after they have undergone
violent relaxation (Aarseth, Lin, \& Papaloizou 1988).

As can be seen from the figure, $\rho_h$ has a fairly shallow radial
dependence, varying approximately as $R^{-0.9}$.  The average density
of the warm phase drops off somewhat more steeply, varying
approximately as $R^{-1.2}$.  The stellar mass density has the
steepest radial dependence, varying approximately as $R^{-3}$.

Figure~2b shows the final radial distributions of the pressure (solid
curve) and temperature (dashed curve) of the hot phase.  At radii near
100~kpc, the temperature remains at $\sim10^6$~K.  At smaller radii,
however, the increasing density of the warm phase, with the
corresponding increase in the heating rate, leads to temperatures
several times higher.  The combination of the density gradient with
the shallow temperature gradient leads to a relatively steep pressure
gradient.  The magnitude of $nT\approx7\times10^{5}$ at 10~kpc and it
varies approximately as $R^{-1}$.  The change in slope outside of
30~kpc is related to the form of the dark matter potential used in the
model, which becomes Keplerian beyond 50~kpc.

Figure~2c shows the radial dependences of the mass ratio of the warm
to the hot phase (solid curve), and of the volume filling factor,
$f_V$ of the WFC's (dashed curve).  We define $f_V$ as
$f_V\equiv\langle\rho_w\rangle/\mu_wn_w$, ie. it is the ratio of the
average density of the warm gas to the internal density of the WFC's.
As can be seen, most of the mass of the system is in the WFC's.  It is
also seen that $f_V$ shows very little variation with radius, and
$f_V\approx0.01$ for $R_c\lapprox R\lapprox100$~kpc.  The lack of
variation in $f_V$ may seem somewhat counterintuitive, given that
$\langle\rho_w\rangle$ decreases at larger radii.  It is a
consequence, however, of the fact that the WFC's are in pressure
balance with the RHG, and the pressure of the latter increases rapidly
towards smaller galactic radii.  The clouds are therefore compressed
into much smaller radii at small $R$.  Note also that the ratio
$\langle\rho_w\rangle / \rho_h > 1$ everywhere, such that most of the
ordinary matter is in the WFC's rather than residing in the RHG.  This
tendency arises because $\tau_c < \tau_d$ initially in this model.

Figure~2d shows the radial dependences of the ratio of the depletion
timescale of the WFC's by star formation to the dynamical timescale
(solid curve), and the ratio of cooling in the RHG by bremsstrahlung
emission to that by conduction into the WFC's (dashed curve).  The
depletion timescale is defined as
\begin{equation}
\tau_{dep}={{\langle\rho_w\rangle}\over{\dot\rho_{sf}}},
\end{equation}
while the dynamical timescale
\begin{equation}
\tau_d={R\over{V_c}}.
\end{equation}
The ratio $\tau_{dep}/\tau_d$ gives a rough estimate of the ability of
the system to replenish warm gas that has been lost to star formation
(it is also replaced by mass lost from the warm phase, but this can be
replenished only by motions at the sound speed of the hot phase,
$c_s\sim V_c$).  As can be seen from the figure, both ratios are
$\gg1$ between $R_c$ and 100~kpc.  Only in the inner regions of the
halo, $\tau_{dep}$ becomes a significant fraction of $\tau_d$ such
that efficient conversion from gas into stars is possible.

For the values of $T_h$ and $\rho_h$ in the model, most of the flux
from bremsstrahlung emission is composed of X-ray photons.  In the low
density outer regions of the halo, bremsstrahlung is not an efficient
radiative process, and a comparable amount of energy is lost to the
RHG via conduction.  Within 100 kpc, the total luminosity of
bremsstrahlung emission is $L_{brem,Tot}=1.2\times10^{11}$~L$_\odot$.
By comparison, the total energy lost from the RHG via conduction into
the WFC's (and immediately radiated from the WFC's) is
$L_{cond,Tot}=1.3\times10^{10}$~L$_\odot$.  In the models discussed
below, a much greater fraction of the energy is lost via conduction.
Based on these results, we do not expect these systems to be luminous
source of X-rays.

\subsection{Model~2, Changes with $V_c$}
We compare the results of the previous model with one having a
circular velocity of 100~km~s$^{-1}$.  Such a model represents a dwarf
system either at any epoch, or one which will become a larger system,
but is at an early stage of evolution and has not yet undergone
significant merging.  In CDM models (Blumenthal et al.
1984; Thoul, \& Weinberg 1996; Navarro, Frenk, \& White 1997; Klypin,
Nolthenius, \& Primack 1997; Abel \& Mo 1998), the first structures
which form are significantly
smaller than this.  However, as can be seen from comparing the 
results below with those of Model~1, systems with significantly
smaller values of V$_c$ are unlikely to contain a significant mass
of RHG, and the assumptions of the models would not be applicable
to such systems.  The results of Model~2 are shown in Figure~3, which
illustrate the same quantities as Figure~2.

In Figure~3a, it can be seen that the densities are reduced
significantly relative to the model with higher $V_c$.  The relative
sense of the steepness is retained, though the slopes are somewhat
different, with $\rho_h$ varying approximately as $R^{-2/3}$,
$\langle\rho_w\rangle$ as $R^{-1.5}$ and $\rho_\ast$ as $R^{-2.5}$.

As shown in Figure~3b, the temperature gradient in the RHG is again
very shallow.  The pressure of the RHG varies approximately as
$R^{-1}$, but it is lower than in the previous model, with $n_h
T_h\approx3\times10^4$ at 10~kpc.

The smaller value of $V_c$ relative to the previous model reduces the
heating efficiency of the WFC's.  In order to balance heating and
cooling, the density of the RHG is reduced relative to the previous
model.  This reduction in the RHG's density relative to that of the
warm phase
is accompanied by increases in the mass ratio of the warm to the hot
phase, and of the filling factor of the warm phase.  Both of these
quantities are larger by an order of magnitude relative to the earlier
model, as can be seen in Figure~3c.  We also note that within 1.5~kpc,
where $f_V>1$, our model assumptions obviously break down.  In this
case, the WFC's should be considered as a coherent entity rather than
an assembly of smaller clouds.

The relatively small pressure of the RHG leads to low internal density
of the WFC's, and a corresponding reduction in the star formation
rate.  This is reflected in the large values of $\tau_{dep}/\tau_d$
shown in Figure~3d.  The reduced density of the RHG and the large
$<\rho_w> / \rho_h$ ratio also increase the relative importance of
conductive cooling of the RHG relative to bremsstrahlung emission.  We
find total luminosities out to 100~kpc of $L_{cond,T}=7.2 \times 
10^8$~L$_\odot$ and $L_{brem,T}=5.1 \times 10^9$~L$_\odot$.  

\subsection{Model~3, Changes with $f$}

In Figure~4, we show the results of a model in which
$V_c=220$~km~s$^{-1}$, and in which $f=0.1$, ie. the WFC's move
inwards with speeds of at most $V_t=0.1V_c$.  This prescription would
represent a situation in which disruption, coalescence, and
coagulation occur on timescales short compared to the acceleration
timescale of the clouds.  Figure~4 shows the same quantities as
Figure~1.

The reduction of the infall velocity has a similar affect as a
relatively small $V_c$, that is, a decrease in the heating efficiency
of the RHG by the WFC's, relative to the $f=1$ case.  Both density and
pressure of the RHG are relatively small in this model, similar to
that seen in Model~2, which has a smaller value of $V_c$.  With a
relatively small density, the RHG loses more energy through conduction
into the WFC's than by bremsstrahlung everywhere in the system (Figure 4d).
The total
luminosities of the two processes are found to be $L_{cond,Tot} = 8.5
\times 10^9$~L$_\odot$ and $L_{brem,Tot} = 1.4 \times 10^8$~L$_\odot$.

The reduction in the infall velocity also leads to a shallower density 
profile in the WFC's.  Comparing with the earlier models, this tendency 
causes star formation to be enhanced at large radii and suppressed at
smaller radii.  Consequently, the stellar density becomes relatively
shallow compared with previous results.

\subsection{Model~4, Domination by small clouds}
We have also considered a model in which $\gamma<-4.5$ so that the
total mass and surface area of the cloud population is dominated by
the small rather than the large clouds.  In this limit, heating of the
RHG, and conductive losses from the RHG into the WFC's are taken from
equations~(24) and (30).  The infall speed of the WFC's is limited to
\begin{equation}
V_t=\left(D_\rho{{S_{es}}\over{R}}\right)^{1\over2}V_c.
\end{equation}

Due to their relatively low infall velocities, individual small clouds
are less efficient at heating the RHG than are large clouds.  For a
given $\langle\rho_w\rangle$, therefore, a distribution dominated by
small clouds leads to a significantly smaller value of $G$.  In
addition, the larger surface-to-volume ratio of a distribution of
WFC's dominated by small clouds leads to a much greater conductive
cooling rate as compared to that in a distribution dominated by large
clouds.

The increase in its efficiency relative to the above models means that
cooling by conduction into the WFC's dominates bremsstrahlung
everywhere in the RHG.  The combined effects of higher cooling
efficiency and lower heating efficiency lead to the depletion of the
RHG and a much larger mass ratio of the WFC's to the RHG relative to
the models above.  The lower density in the RHG relative to earlier
models leads directly to smaller pressures, causing a reduction in
both $n_w$ and
the star formation rate.  The change was so dramatic that we found
$f_V>1$, ie. the RHG effectively does not exist in the system. But, in
the absence of RHG, the WFC's cannot fragment through Kelvin-Helmholz
instability to offset their growth through cohesive collisions (see
\S3.3). Thus, the large $<\rho_w > / \rho_h$ ratio and the dominance
of the small clouds appears to be incompatible. A self-consistent
treatment of both phases, including the size distribution of the WFC's
will be examined and discussed in a future paper.

The above models assume that $\Lambda_w=10^{-22}$ergs~cm$^{3}$~s$^{-1}$,
a high value for warm clouds, near that of the peak from hydrogen 
emission.  Such would be expected, given the stability arguments
presented in \S~3, if the small cloud population is dominated by
those clouds which are just able to radiate away the energy
conducted into them from the RHG.  If, however, the small cloud
population is dominated by slightly larger clouds, which can radiate
more efficiently, then a much small value of $\Lambda_w$ may be
more appropriate.  We have therefore repeated Model~4 using
$\Lambda_w=10^{-24}$ergs~cm$^{3}$~s$^{-1}$.  The decreased
emission efficiency of the small clouds led to results that were
significantly different from those found above, and closely 
resembled those found for Model~3.  The difference between these two
possible cases remains an uncertainty in this work, and shall be
resolved in upcoming models which examine the structure of WFC's
embedded within the RHG.

\section{Discussion}
In this paper, we examine the microphysics of gas dynamics in the
early epoch of galactic evolution.  Our objective is to provide a
description of the dominant physical processes in ordinary matter
which may be applied to a large class of galaxy formation models. For
example, in the canonical hierarchical galaxy formation scenario,
small dwarf-galaxy building blocks, containing non-interactive dark
matter and gaseous ordinary matter, form first and subsequently merge
to form larger entities such as our Galaxy.  In the present analysis,
we neglect the dynamical evolution of the dark matter halo which is
undoubtedly important in determining not only the formation process
but also the present kinematic properties of galaxies.  This
approximation enables us to focus our attention upon the evolution of
gas in the early epoch of galactic evolution, which regulates the rate
and location of star formation and therefore the light distribution
and chemical properties of the emerging galaxies.  We discuss below
three possible applications of our results.

\subsection{Galactic Stellar halo}
The results presented here show that thermal instability results in
the formation of the residual halo gas (RHG), with density and pressure
appropriate for quasi-hydrostatic and energy equilibria, and warm
fragmentary clouds (WFC's), which are
pressure confined by RHG and heated by ionizing UV photons.  Such
systems have been proposed as being the source of observed
Lyman-limit systems (Mo \& Miralda-Escude 1996).  The mass limits of
the WFC's are set by the same criteria as in the earlier work, but
in this work we include many more details of the interactions
between the phases.

Inside
$\sim 10-100$~kpc, WFC's in galaxies with masses comparable
to the Galaxy are self-shielded from the extragalactic UV flux (see
Figure 1). Unless these WFC's are continually heated by the UV flux
from nearby massive stars, further cooling reduces their temperature
to $\leq 10^2$ K.  Gravitational instability in the large WFC's
leads to spontaneous formation of stars among which the massive stars
radiate UV photons, ionize their surroundings, and quench the
formation of additional stars during their lifetime (Lin \& Murray
1992).  Through such a self-regulating feedback process, the maximum
rate at which gas may be converted into stars is determined by the
maintenance of an adequate UV flux to photoionize all the WFC's
(\S4.2).

In the regions far from the center of the halo where the density of
the RHG is relatively low, the internal density of the WFC's and their
average density $<\rho_w>$ are also low.  The effect of self-regulation
limits the star formation timescale $\langle\rho_w\rangle /
\dot{\rho}_{SF} > \tau_d$.  But, the star formation efficiency is much
higher at smaller galactic distances.  The results in Figures 2-4 show
that within a few kpc (depending on the model), the stellar density
$\rho_\ast$ already exceeds both $<\rho_w>$ and $\rho_h$ after 1 Gyr.
Once the stars are formed out of gas, they cannot dissipate their 
orbital energy such that their orbital radii cannot contract further.
Thus, the interaction between RHG, WFC's, and the newly formed massive
stars essentially determines the asymptotic surface brightness distribution
in galaxies.  In Figure~5, we show the computed surface density profiles
computed from the stellar distributions of Models~1-3.  As can be seen,
the models are fairly well fit by deVaucouleur profiles within the
inner 50~kpc, in good agreement with the profiles of spheroidal systems.

The onset of rapid and efficient star formation invalidates the
instantaneous mixing and incremental gas-to-star conversion
assumptions which are essential to the closed-box models for galactic
chemical evolution and enrichment (cf Binney \& Tremaine 1987),
despite the apparent consistency between it and the observed
metalicity distribution among population II stars (Chiappini et al. 1999).
If these stars are formed in a series of starburst events in
which a large fraction of the remaining gas is converted into stars on
a timescale shorter than the dynamical timescale of the galaxy or the
lifespan of the massive stars, the metallicity distribution would
reflect the metallicity inhomogeneity in WFC's.  In this case, the
deficiency of extremely metal poor stars (with [Fe/H]$\lapprox$-3) would
be consistent with an evolving initial mass function which gradually
becomes less biased towards massive stars as the WFC's are chemically
enriched.

A necessary condition for the formation of stars with long-lived, 
low-mass stars is small $M_{crit}$ (see \S4.2).  In a self-regulated
environment, WFC's with [Fe/H]$\gapprox$-3 can spontaneously cool
from $10^4$ K to $\sim 10$ K (so that $M_{crit}\lapprox1M_\odot$) between
successive generations of nearby massive stars (with an interval
$\sim$ a few $10^6$ yr) provided the initial $n_w > 0.1-1$
(Hellsten \& Lin 2000).  The corresponding external pressure needed 
to confine such WFC's is $nT\sim 10^{3-4}$.  In Figures~2-4, we see that
these
values of $n_h T_h$ are attained outside 10~kpc, similar to the regions
where the metal poor population II stars are located.

In the above scenario, we have neglected the effect of magnetic fields
which stablize cold clouds against gravitational instability through
field-ion coupling and ion-neutral collisions (cf Shu 1985; Shu, Adams,
\& Lizano 1993).  In dense
cores of molecular clouds around the solar neighborhood, low-level
ionization is maintained through cosmic ray heating and the field
strength declines through ambipolar diffusion (Spitzer 1978; Shu 1985).
If this process is important in WFC's, the star formation rate would
be much reduced from those illustrated in Figures~2-4.  Collisions
between WFC's would occur at velocities in excess of 10~km~s$^{-1}$.
Shock compression and rapid cooling near the collision interface could
lead to a rapid expulsion of magnetic field, rendering its support
ineffective.  In the models discussed above, the covering factor of
large WFC's is of order unity.  A large cloud would therefore collide
with another large cloud only about once during a galactic crossing
time.
Collisions between large and small clouds would, however, occur much
more frequently.  In the limit that the size distribution of the
clouds is dominated by relatively small clouds, the area covering
factor is large, and small clouds collide with each other much
more frequently.  These issues are beyond the scope of the present
investigation and they need to be thoroughly investigated in the
future.

\subsection{Formation of globular clusters}
The analysis presented here can also be applied to the formation of
Galactic globular clusters (Lin \& Murray 1996). Prior to the
conversion of ordinary matter from gas into stars, the progenitors of
these clusters were protocluster clouds (PCC's).  The chemical
homogeneities within individual clusters and the large metallicity
variation among different clusters suggest that PCC's are a
distinct entities which must be confined either by their own self-gravity
or external pressure. But, if PCC's are entirely bound by their
own self-gravity, external UV heating would not be adequate to
suppress thermal instability within them (see \S3.3).

The magnitude of $M_{crit}$ for the WFC's is comparable to the mass of
globular clusters.  We identify these warm, marginally
self-gravitating and partially pressure-confined WFC's as PCC's.  At
Galactic distances $D \sim 3-30$ kpc, PCC's with mass ($M$) $ <
M_{crit} (\sim 10^6 M_\odot)$ are confined by the the pressure of the
RHG, $n_h T_h \sim 10^{3-5}$, depending on the Galactic halo structure
during the epoch of cluster formation (Figs. 2-4).  If these PCC's are
completely ionized and have a $T_w \sim 10^4$, then $n_w \sim
0.1-10$~cm$^{-3}$.  For these structural parameters, the UV flux
needed is equivalent to that emitted by a few O5 stars at a distance
comparable to or greater than their size (typical a few pc) (see
Figure~1). (These stars could also reside within the PCC's).  These
clouds could persist for a significant fraction of $\tau_d$ if the
accretion of smaller clouds or condensation from the RHG is adequate to
compensate for their mass loss due to stripping by the RHG.

On the observational side, in order to verify that the PCC's were
pressure-confined, we first estimate their $n_w T_w$ from the current
properties of globular clusters, averaged over their half-mass radii
($r_h$) (Murray \& Lin 1996).  We use these quantities because the
stellar density and the velocity dispersion at $r_h$ do not change
significantly during post-formation evolution.  Extrapolation to the
stage prior to star formation is, however, highly uncertain. If, after
their formation, the young stellar objects undergo collapse and
virialization from rest, the clouds' initial radii ($r_i$) would be
$\sim 2 r_h$.  Larger $r_i$ would be expected if star formation
requires dissipative collisions and coagulation of substellar
fragments (Murray \& Lin 1996). But $r_i$ is unlikely to be larger
than the tidal radii of the PCC's, which are typically a few times
larger than $r_h$.  Thus, the initial density of the PCC's may be 1-3
orders of magnitude smaller than the average cluster density at $r_h$
today.  Based on the present velocity dispersion of the clusters, we
infer the initial temperature of the PCC's to be $\sim 10^4$~K,
comparable to that expected if they were photoionized.  From these
estimates, we infer $n_w T_w \sim 10^{2-5}$ (Murray \& Lin 1992). The
dependence of the pressure upon galactic radius, $D$, is very poorly
determined from
the observational parameters.  Of more significance is that the
magnitude of the pressure inferred from the observations is very
similar to that found in the RHG in our models (Figs. 2-4).

From these results, and the
cluster metallicities, we can also estimate the cooling time scale
($\tau_{cc}$) and dynamical time scale ($\tau_{dc}$), of the PCC's.
The ratio $\tau_{cc}/\tau_{cd}$ increases from $\sim 10^{-4}$ near the
Galactic bulge to $\sim 1$ at $\sim 100$ kpc.  In most PCC's,
$\tau_{cc} < < \tau_{cd}$ and thermal equilibrium is only possible in
the presence of external UV photons with a flux comparable to that
required by self-regulated star formation in the halo.

\subsection{X-ray Luminosity}
In the hierarchical galaxy formation process, coalescence of small
stellar systems (dwarf galaxies) occurs within a few dynamical
timescales.  If the ordinary matter contained initially within these
building blocks is to be heated through shock dissipation to the
virial temperature of the common halo, the mergers would become
luminous X-ray sources (Eke, Navarro, \& Frenk 1998).  But the results
of our models
indicate that; 1) thermal instability leads to the formation of WFC's
which contain most of the ordinary matter, and 2) even in the RHG,
conduction may be a more efficient channel of heat loss than
bremsstrahlung emission.  As discussed in \S~5.2-5.4, the resulting
x-ray luminosities are small, ranging from 10$^{11}$~L$_\odot$ in
Model~1 to 10$^{8}$~L$_\odot$ in Model~3. The spectra of the models
are shown in Figure~6.  The spectra of the low luminosity systems may
be modified somewhat by supernova emission.  Using the total star
formation rates of Models~1-3, we estimate supernova luminosities of
10$^{10}$~L$_\odot$ for Model~1, 6$\times$10$^{9}$~L$_\odot$ for
Model~2, and 5$\times$10$^{8}$~L$_\odot$ for Model~3.  Based upon
these results, we expect relatively little X-ray luminosity to be
released from regions with ongoing galaxy merging events, 
consistent with x-ray observations (Fabbiano \& Schweizer 1995; 
Read \& Ponman 1998).

\subsection{Limitations}
In our first attempt to investigate the complex physics of multi phase
gas dynamics during the early epoch of galaxy formation, we have
adopted various simplifying assumptions such as 1-D spherical symmetry
and power-law size distribution for WFC's. These treatments can be
improved with a self-consistent analysis of the WFC's evolution which
will be presented in a follow up paper.  We have also neglected the
evolution of dark matter which dominates the potential. The results in
Figs 2-4 show that the distribution of pressure, density, filling
factor of WFC's and the star formation rate depend, though not
sensitively on the potential.  A full study of this problem will 
require the use of multi-dimensional integrated simulations of both
dark and ordinary matter.  While such work has been done in a 
cosmological context, the extreme temperature range spanned by the
gas within an individual galaxy (at least four orders of magnitude)
places severe demands upon a code to be able to resolve structures
over many orders of magnitude in size.  Our basic approach and the
prescription provided here can be readily used in such investigations
as they begin to be made.

\acknowledgements
We thank Drs. A. Burkert, S. Faber, U. Hellsten, C. Frenk and A.
Wolfe for valuable conversations.  This research has been supported in
part by the NSF through grant AST-9618548 and by NASA through an
astrophysics theory program grant which supports a joint Center for
Star Formation Studies at NASA-Ames Research Center, UC Berkeley,
and UC Santa Cruz.  This work was performed in part under the
auspices of the U.S. Department of Energy by the University of
California Lawrence Livermore National Laboratory under contract
No. W-7405-Eng-48.

\clearpage

\begin{figure}
\clearpage
\plotone{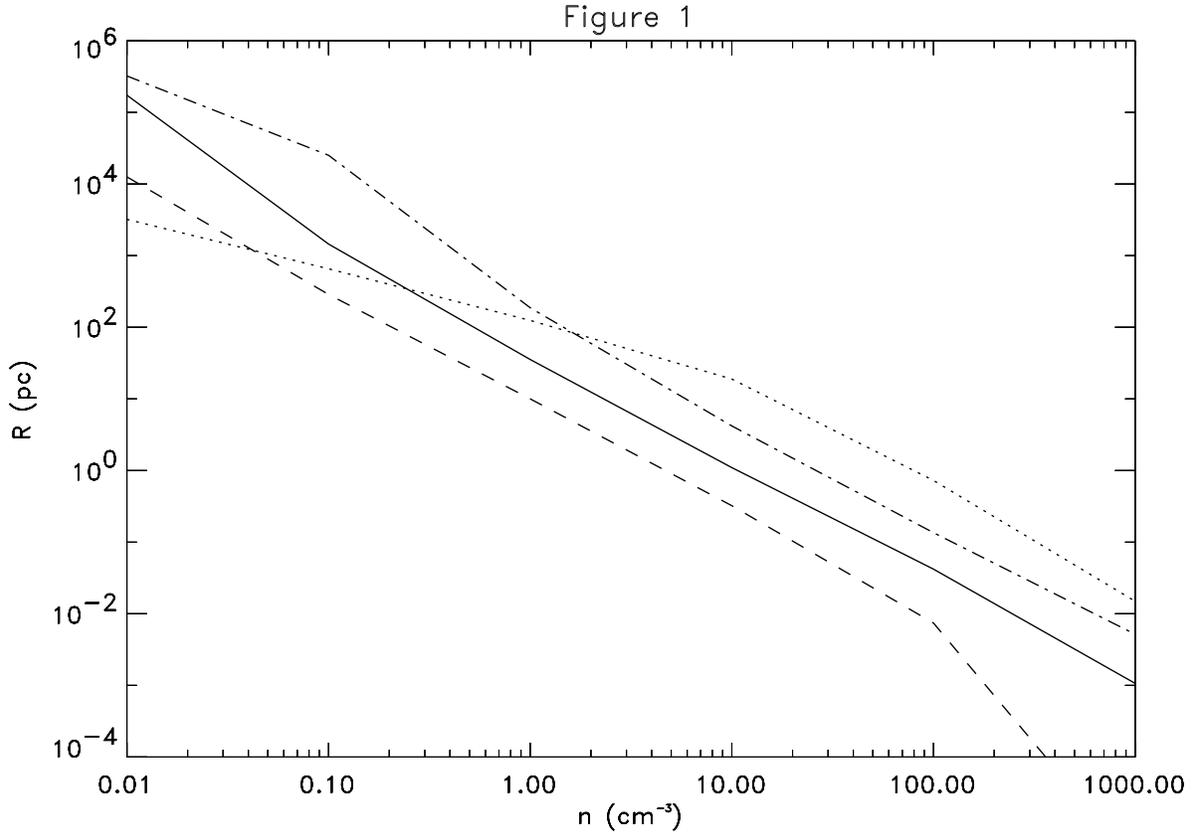}
\caption{The depth into which a cloud may be ionized by various UV sources,
calculated using CLOUDY.  The solid, dashed, and dot-dashed curves are
computed using the standard AGN spectrum of CLOUDY.  The solid curve 
corresponds to
a flux of $0.5\times10^{-21}{\rm\ ergs}{\rm\ cm}^{-2}{\rm\ s}^{-1}
{\rm\ Hz}^{-1}{\rm\ Sr}^{-1}$ at 13.6~eV, while the dot-dashed and dashed
curves correspond, respectively, to fluxes larger and smaller by an order
of magnitude.  The dotted curve is computed using a Kurucz spectrum, with
the UV flux set to correspond to that of an O5 star at a distance of 10~pc.}
\end{figure}

\begin{figure}
\clearpage
\plotone{fig2a.epsi}
\end{figure}

\begin{figure}
\clearpage
\plotone{fig2b.epsi}
\end{figure}

\begin{figure}
\clearpage
\plotone{fig2c.epsi}
\end{figure}

\begin{figure}
\clearpage
\plotone{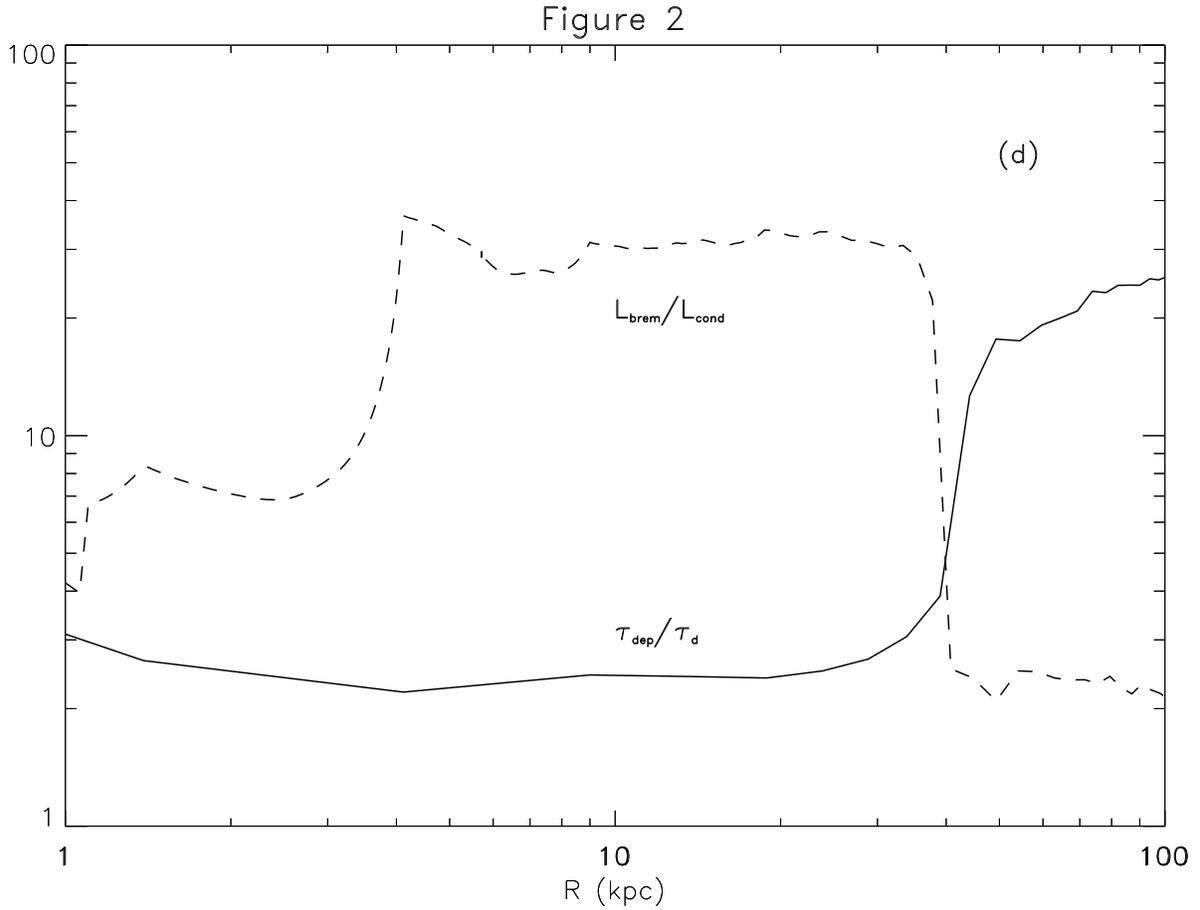}
\caption{Model~1 at 10$^9$~yr.  In (a) are shown the 
density of hot gas (solid), the average density of the warm gas (dashed), and
the stellar mass density (dot-dashed).  In (b) are shown 
the temperature (dashed) and pressure (solid) of the hot phase.  In (c)
are shown the mass ratio of the warm and hot phases
(solid) and the volume filling factor of the warm clouds (dashed).  In (d) are
shown the ratio of the depletion timescale to the dynamical timescale (solid)
and the ratio of the efficiency of cooling of the hot phase by bremsstrahlung
emission to that by conduction into the cool clouds (dashed).  All quantities
are shown as functions of galactic radius.
}
\end{figure}

\begin{figure}
\clearpage
\plotone{fig3a.epsi}
\end{figure}

\begin{figure}
\clearpage
\plotone{fig3b.epsi}
\end{figure}

\begin{figure}
\clearpage
\plotone{fig3c.epsi}
\end{figure}

\begin{figure}
\clearpage
\plotone{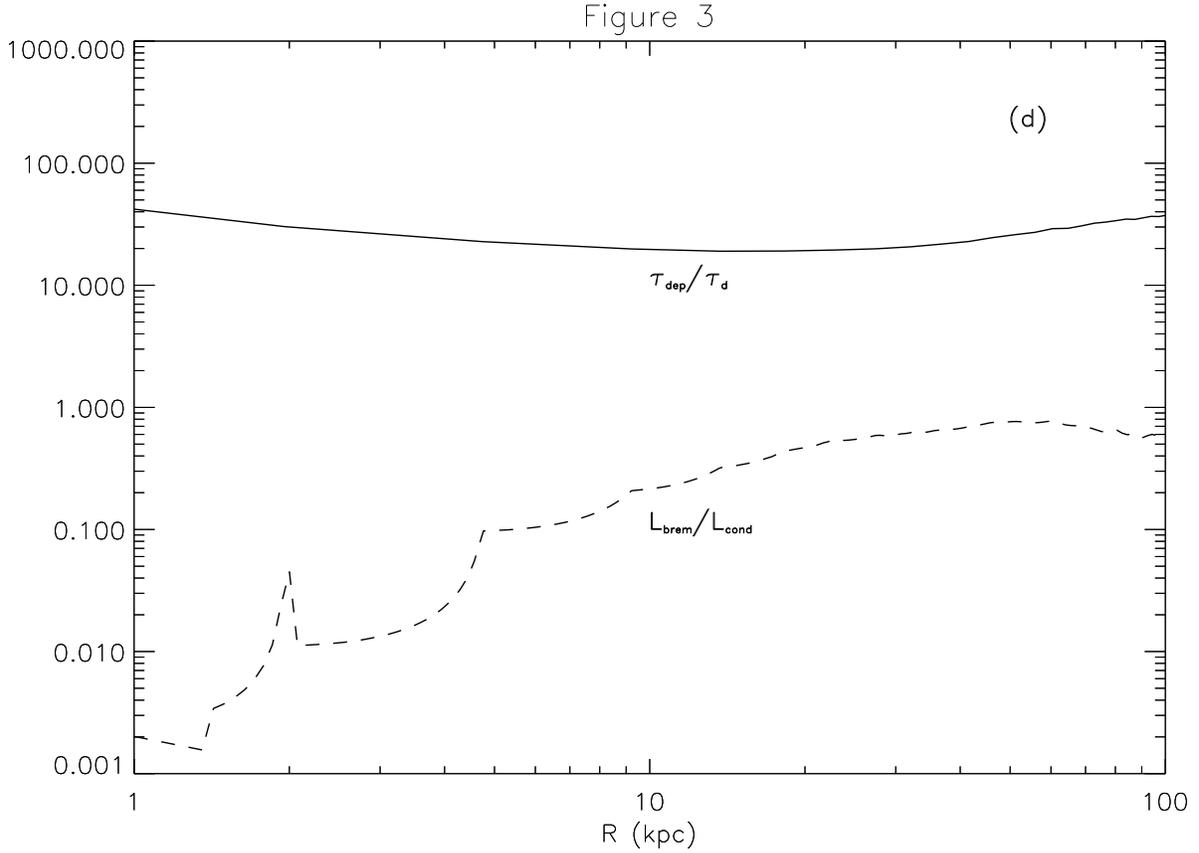}
\caption{Model~2 with $V_c=100$~km~s$^{-1}$ at 10$^9$~yr, displayed as in
Figure~2.  With a lower circular speed, the heating of the RHG by motions of
the WFC's is reduced relative to that of Model~1.  As compared to Model~1,
Model~2 has similar mean density of the
warm gas, whereas the density of the RHG is reduced by approximately an order
of magnitude.  The filling factor of the warm clouds is thus about an order
of magnitude greater than in Model~1.  The reduced pressures relative to
Model~1 lead to lower density within the
WFC's, reducing the star formation rate necessary to maintain the warm phase,
and so the stellar density is also reduced relative to Model~1.  The low
density of the RHG relative to Model~1 also leads to conduction into the
WFC's losses playing a dominant role in energy loss from the RHG, contrary
to the case in Model~1.
}
\end{figure}

\begin{figure}
\clearpage
\plotone{fig4a.epsi}
\end{figure}

\begin{figure}
\clearpage
\plotone{fig4b.epsi}
\end{figure}

\begin{figure}
\clearpage
\plotone{fig4c.epsi}
\end{figure}

\begin{figure}
\clearpage
\plotone{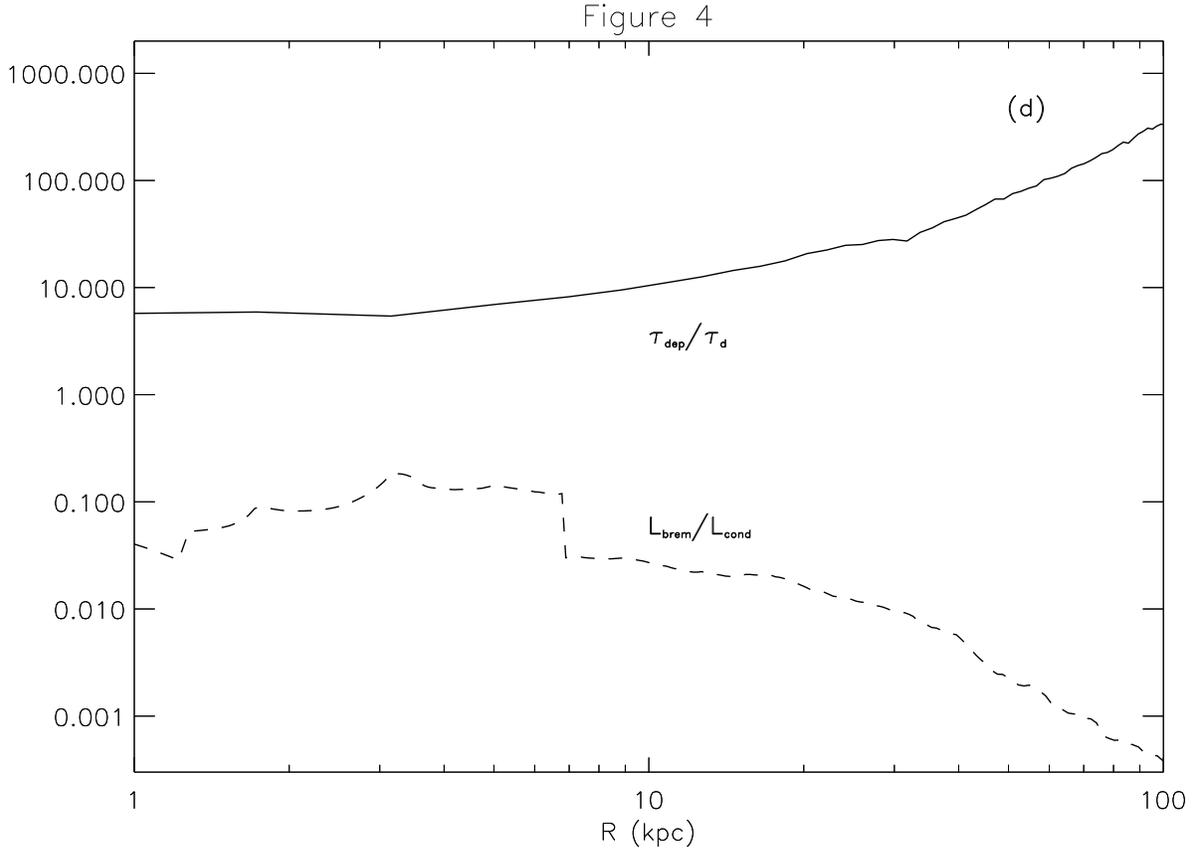}
\caption{Model~3, with $f=0.1$, at 10$^9$~yr, displayed as in Figure~2.  The
effect of $f<1$ is to reduce the heating of the RHG by motion of the WFC, and
the comparison with Model~1 is similar to that of Model~2.  Relative to 
Model~1, the mean density of the warm gas in Model~3 is comparable, whereas
the density of the RHG is lower by an order of magnitude at small radii, and 
more at large radii.  The filling factor of the warm clouds is therefore at
least an order of magnitude larger in Model~3 than in Model~1, and reaches
unity at large radii.  The reduced pressure of the RHG relative to Model~1
leads to reduced densities in the WFC's and therefore a reduced star
formation rate.  The stellar density of Model~3 is therefore lower by an
order of magnitude relative to Model~1.  The lower density of the RHG also
leads to conductive losses dominating bremsstrahlung emission in the hot
phase at all radii.
}
\end{figure}

\begin{figure}
\clearpage
\plotone{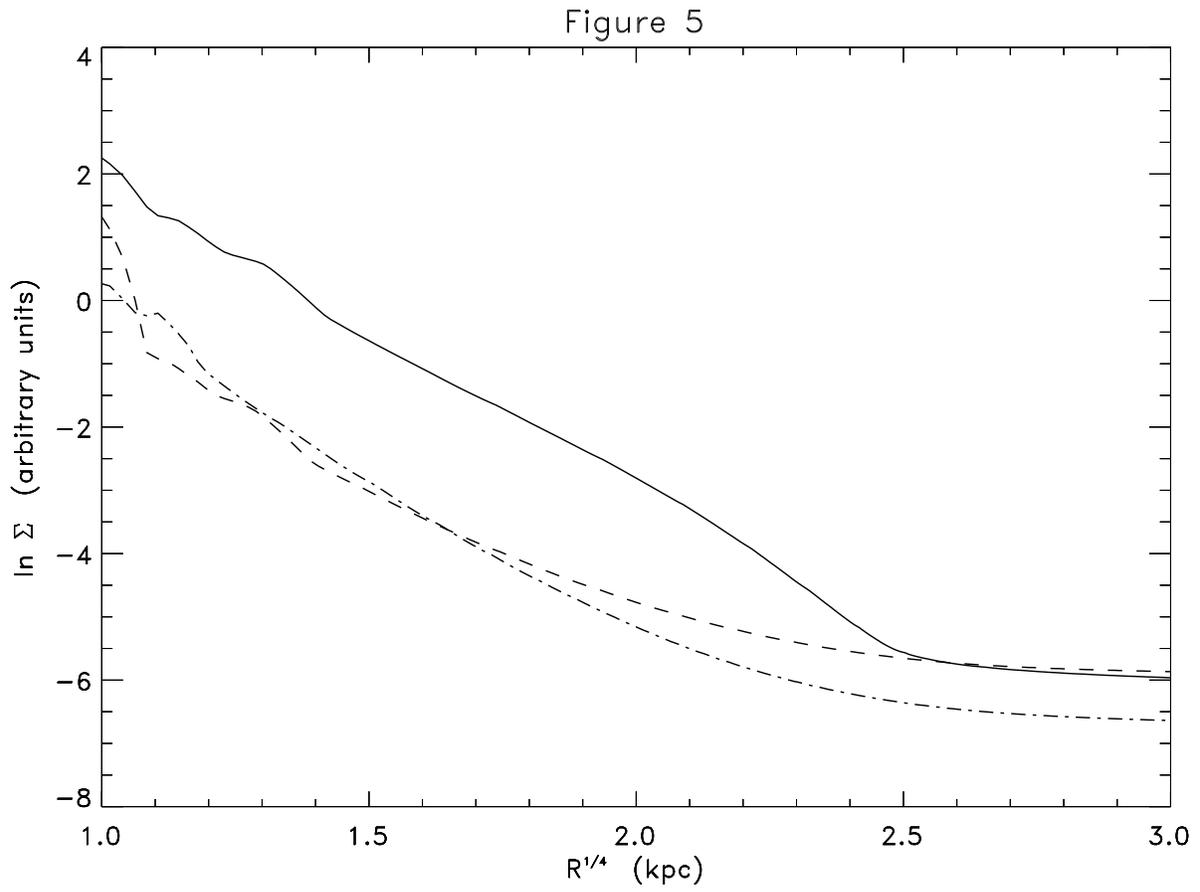}
\caption{The project stellar surface densities of Model~1 (solid),
Model~2 (dashed), and Model~3 (dot-dashed).
}
\end{figure}

\begin{figure}
\clearpage
\plotone{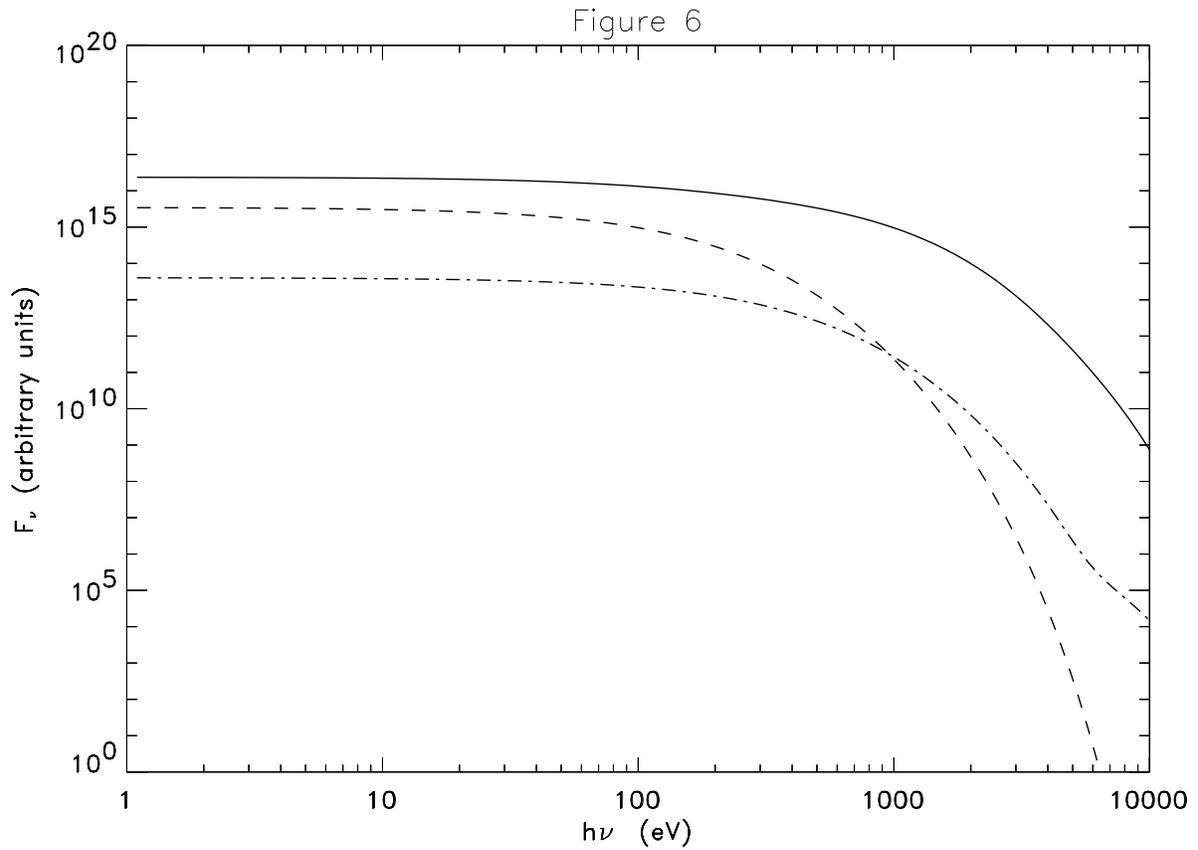}
\caption{The spectral energy distributions (in arbitrary units) of
Model~1 (solid), Model~2 (dashed) and Model~3 (dot-dashed), as 
determined by assuming bremsstrahlung emission at the temperature of
the RHG.
}
\end{figure}

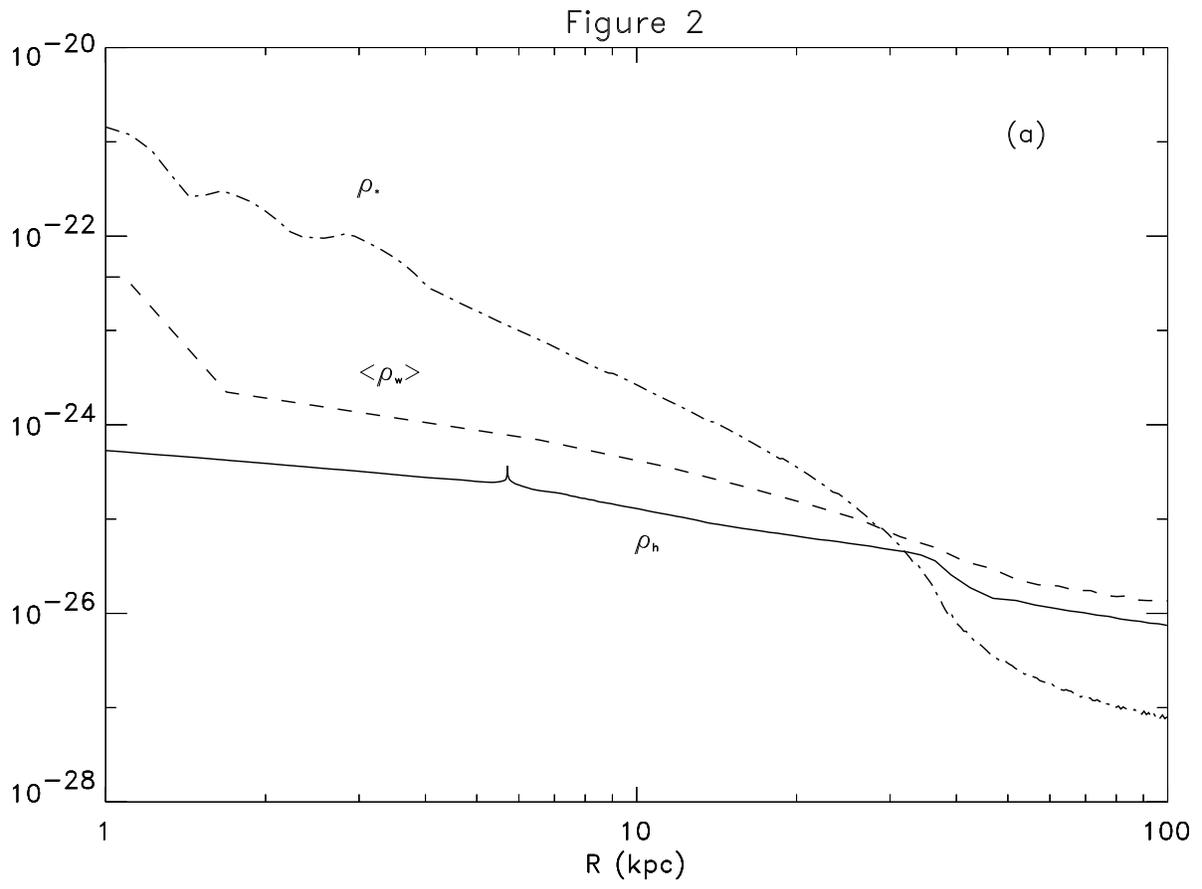
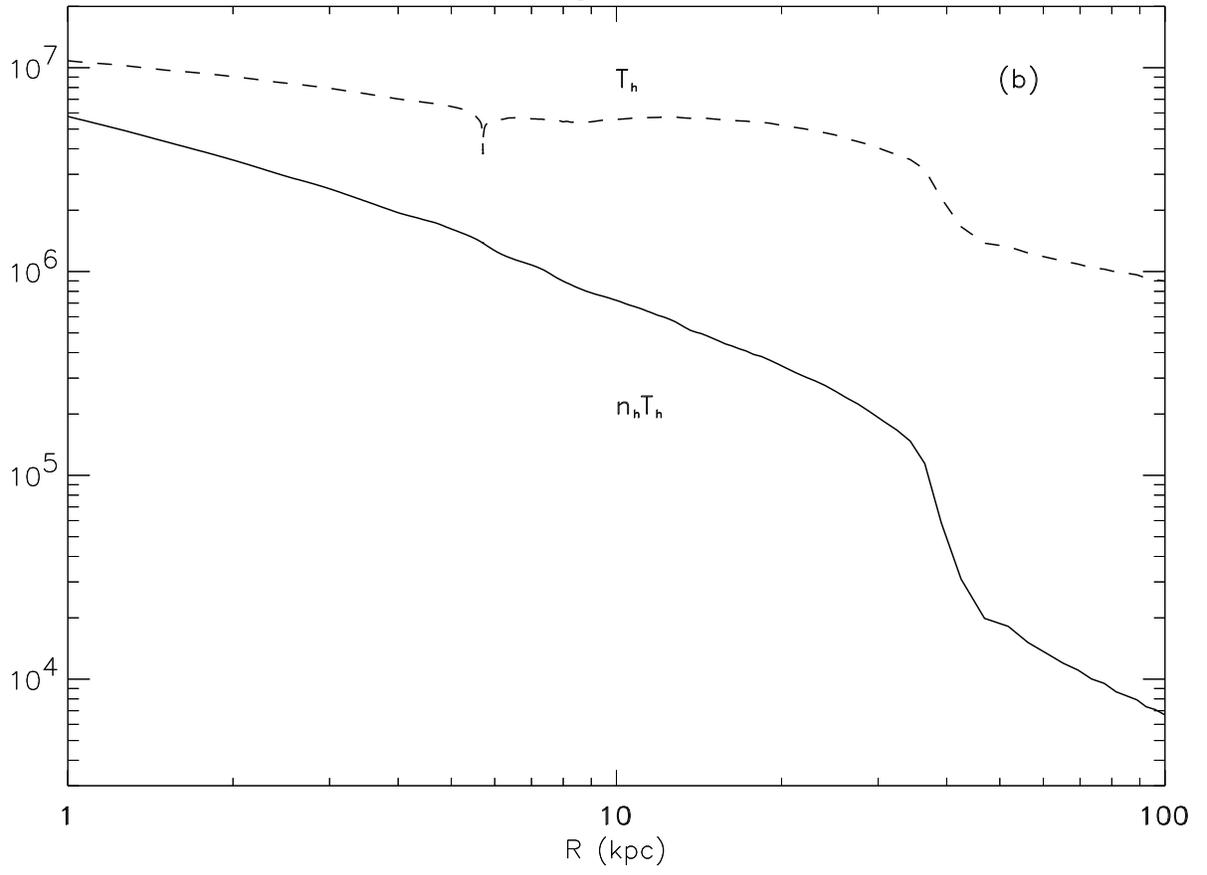
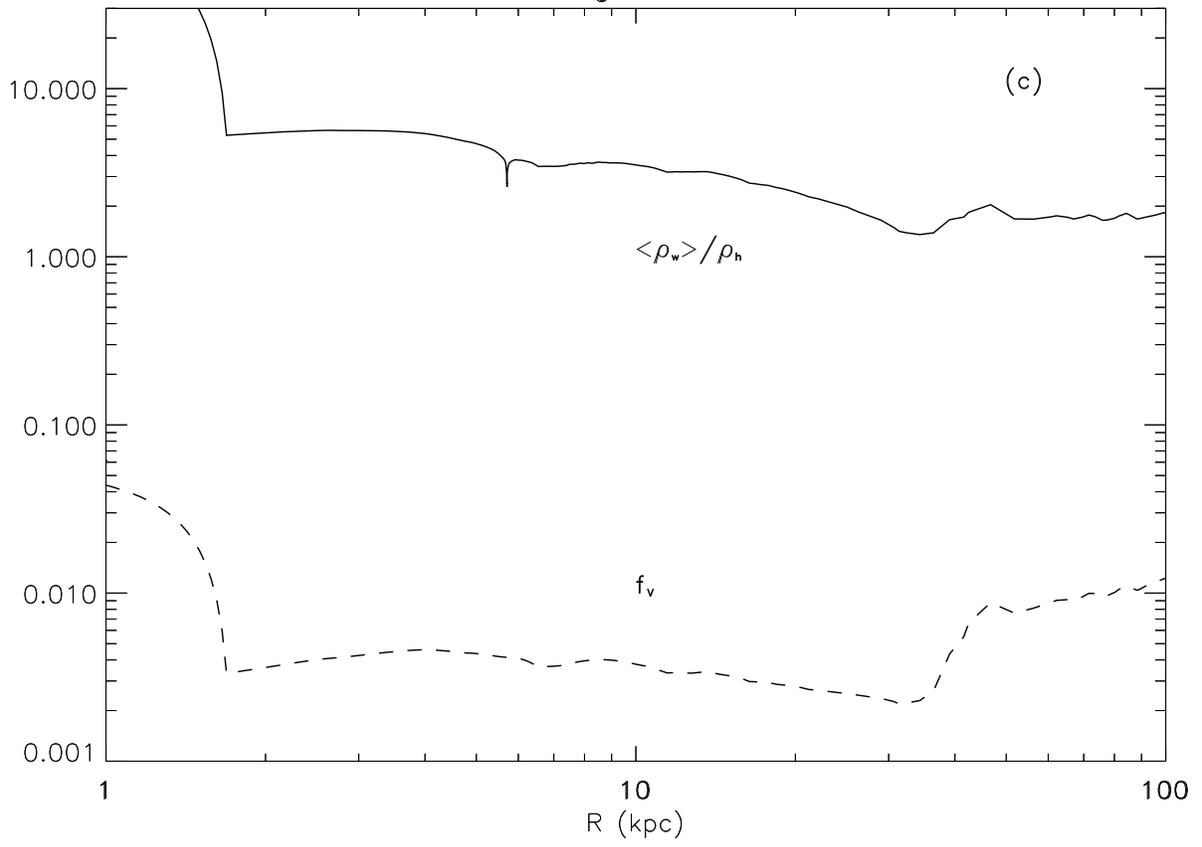
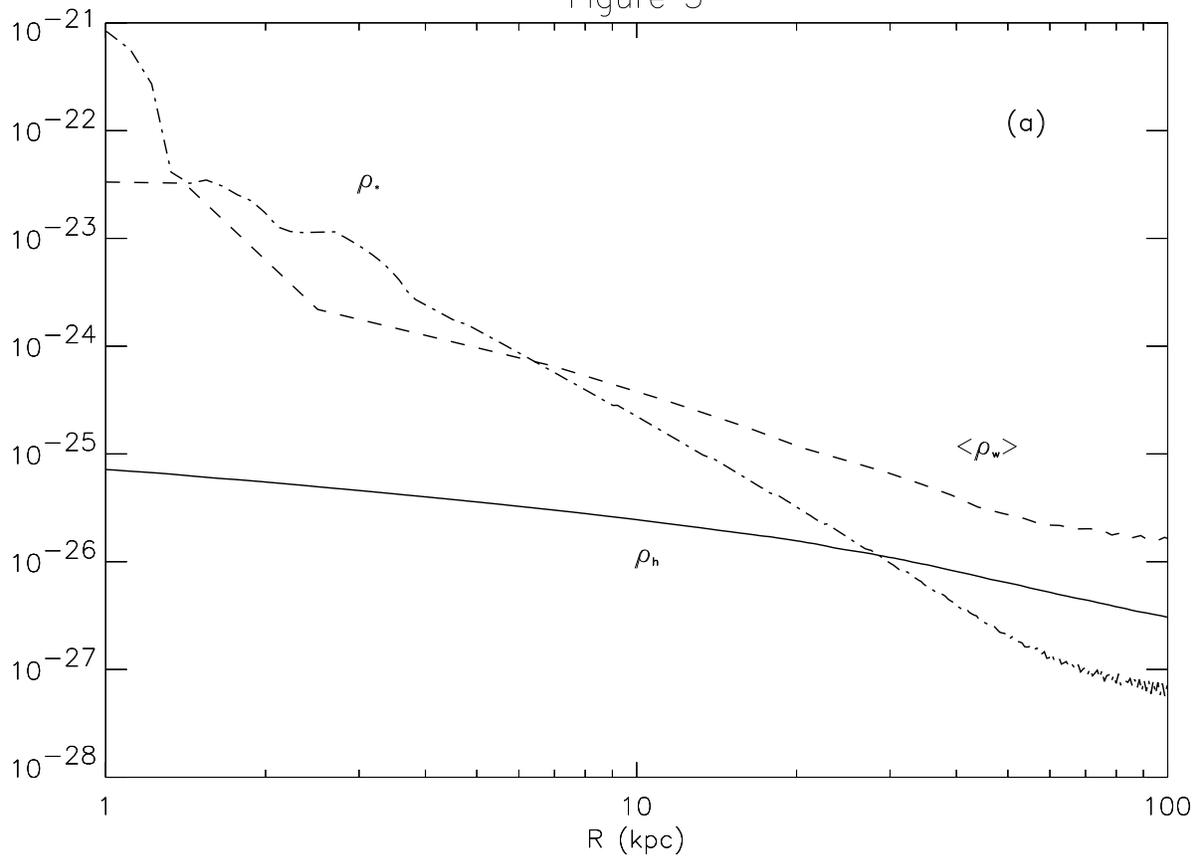
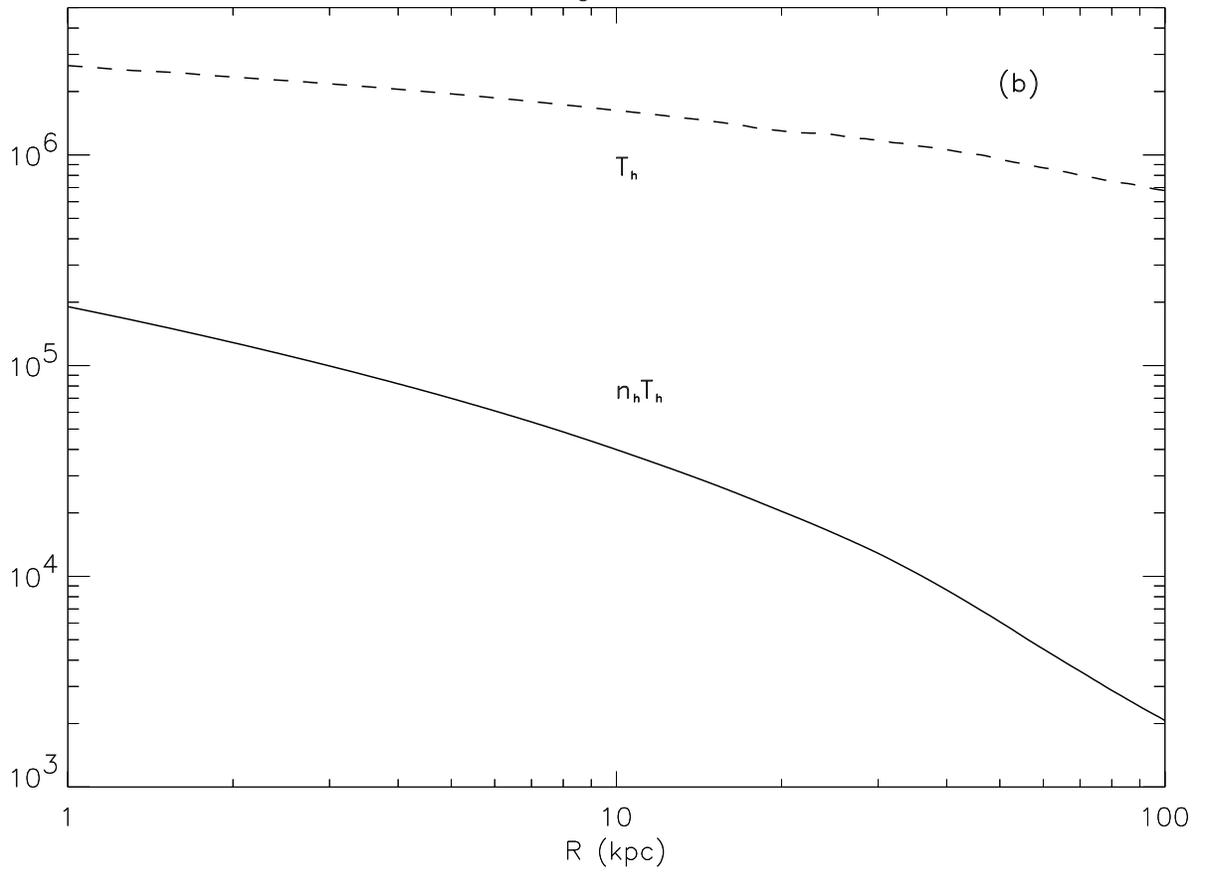
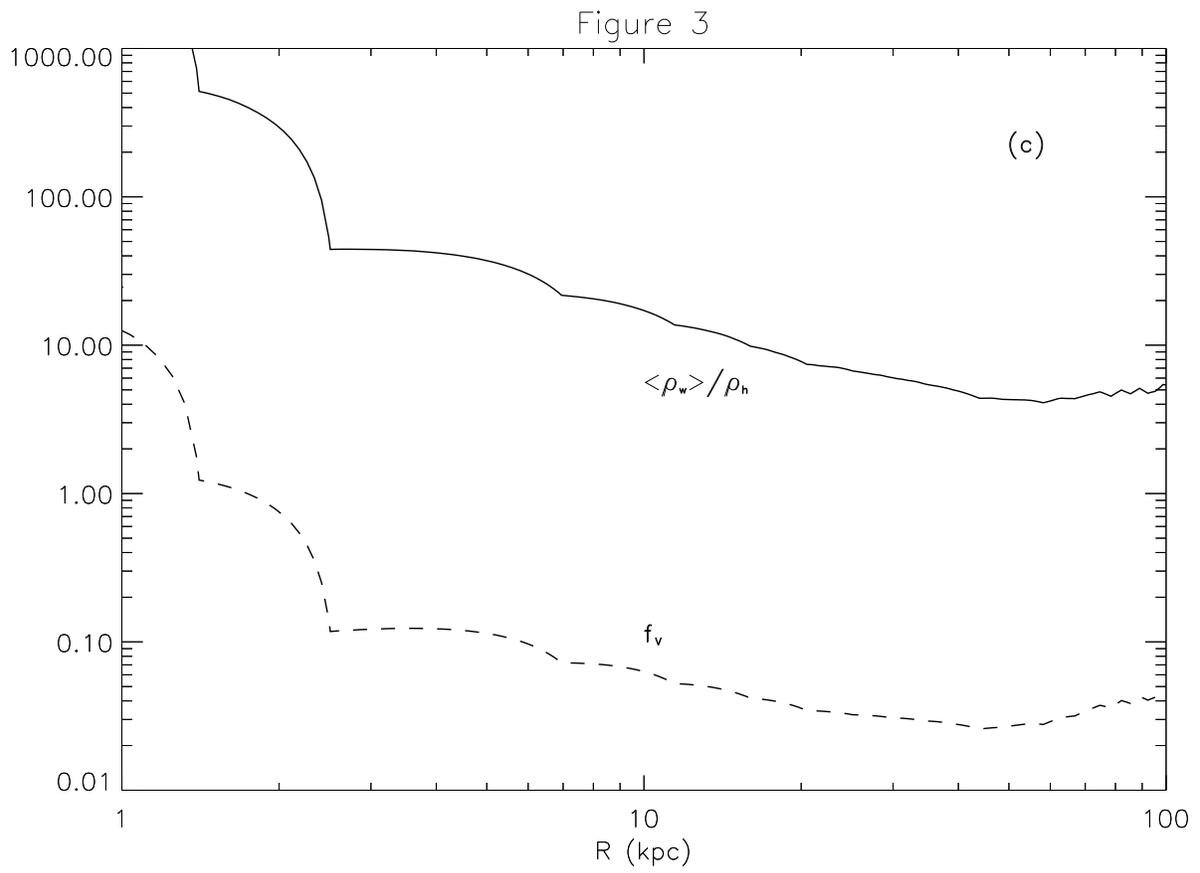
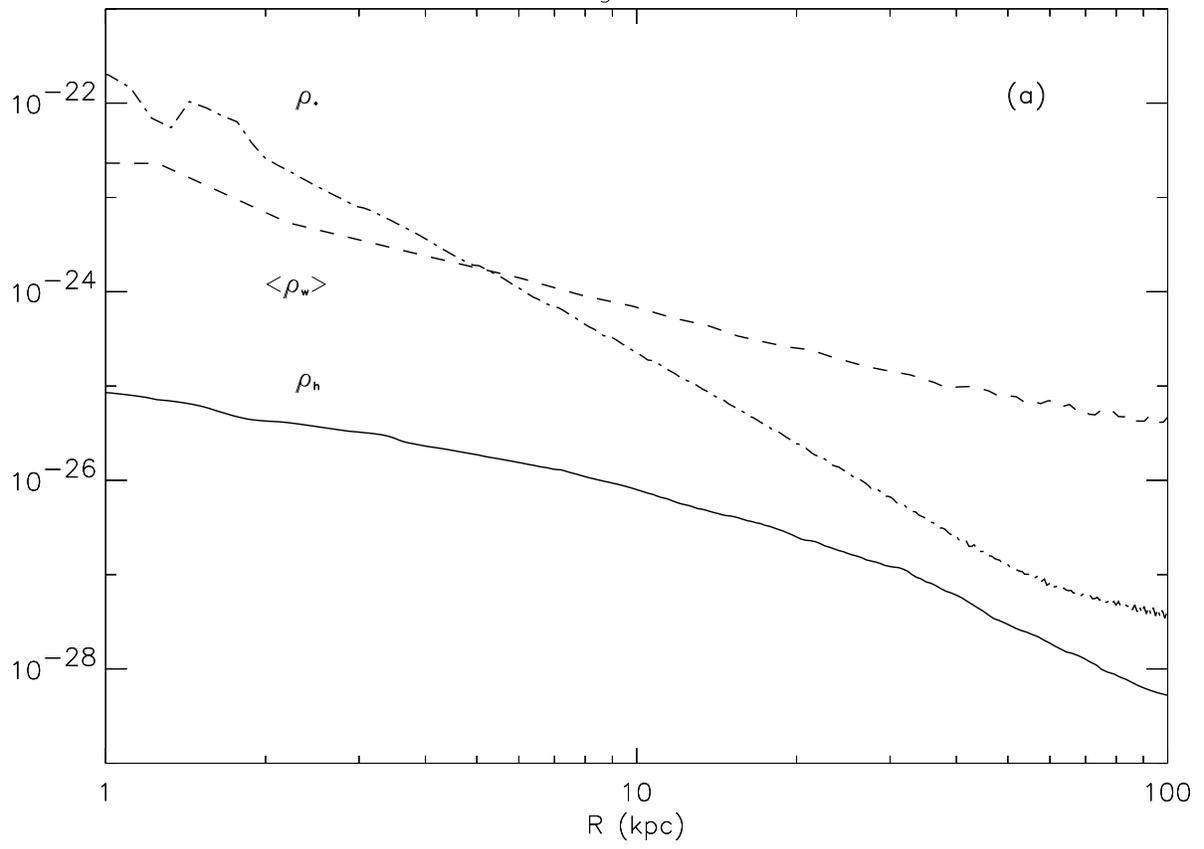
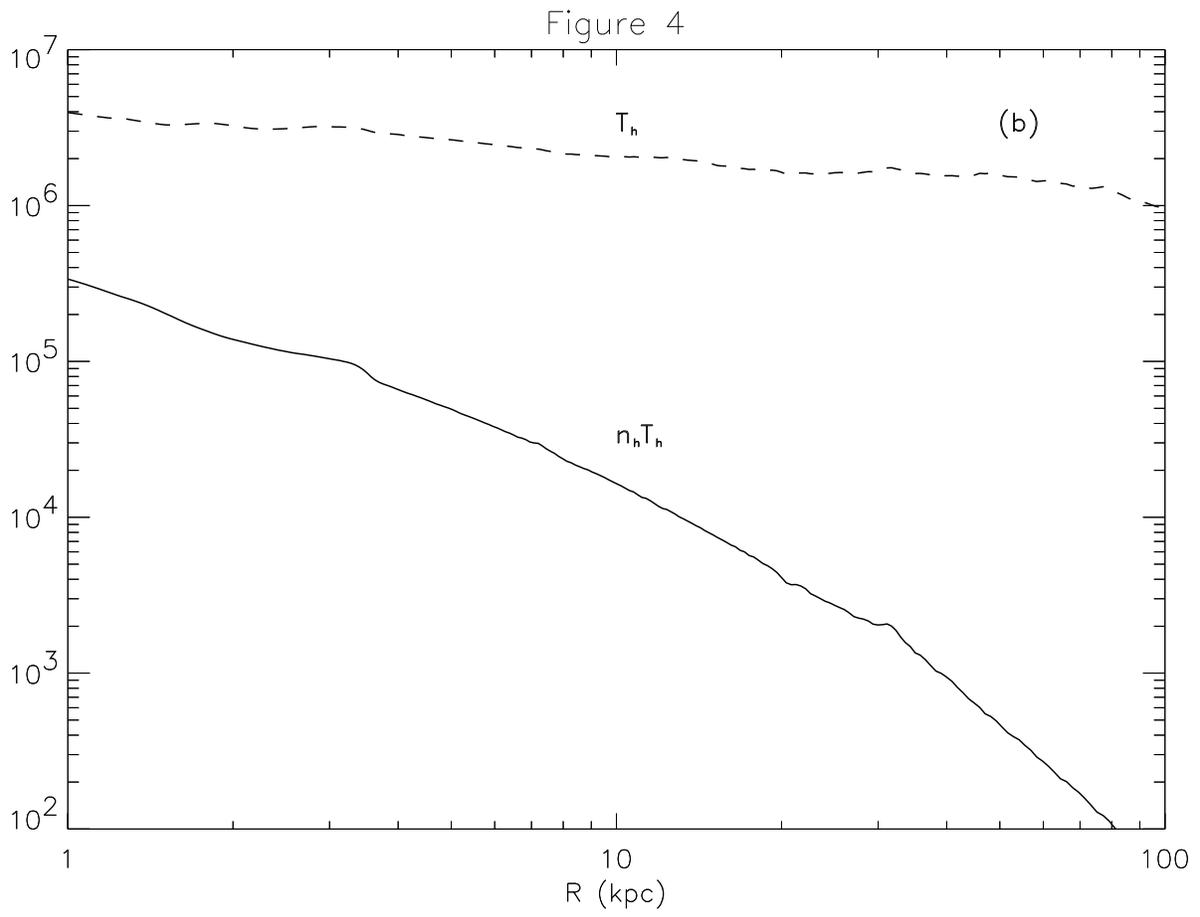
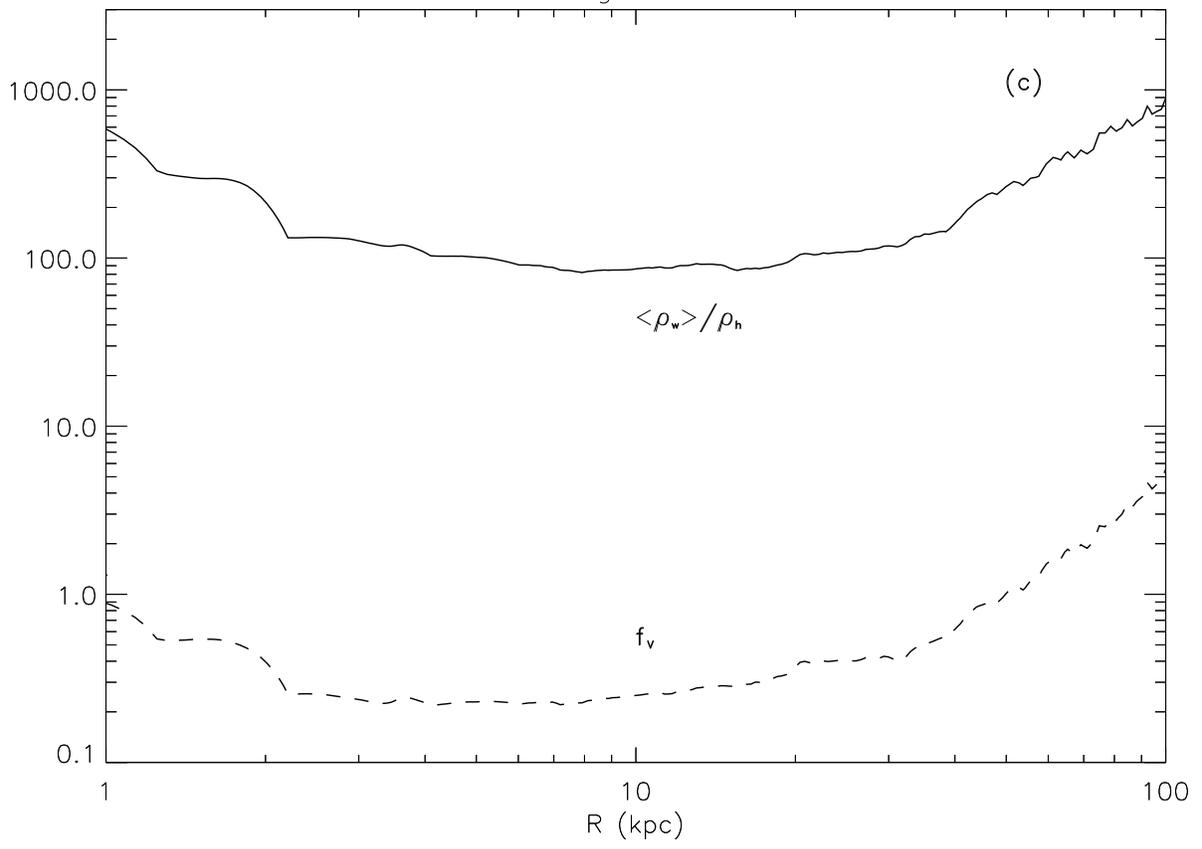
\end{document}